\documentclass[10pt,letterpaper]{article}
\usepackage[top=0.85in,left=2.75in,footskip=0.75in]{geometry}

\usepackage{amsmath,amssymb}
\usepackage{soul}

\usepackage{changepage}

\usepackage[utf8x]{inputenc}

\usepackage{textcomp,marvosym}

\usepackage{cite}

\usepackage{nameref,hyperref}

\usepackage{microtype}
\DisableLigatures[f]{encoding = *, family = * }

\usepackage[table]{xcolor}

\usepackage{array}

\newcolumntype{+}{!{\vrule width 2pt}}

\newlength\savedwidth

\usepackage{setspace} 
\raggedright
\usepackage[aboveskip=1pt,labelfont=bf,labelsep=period,justification=raggedright,singlelinecheck=off]{caption}

\def\done[#1]{{\color{red}{#1}}}
\def\orad[#1]{{\color{olive}{#1}}}

\makeatletter
\renewcommand{\@biblabel}[1]{\quad#1.}
\makeatother

\usepackage{lastpage,fancyhdr,graphicx}

\fancyheadoffset[L]{2.25in}
\fancyfootoffset[L]{2.25in}
\begin{document}
\vspace*{0.2in}

\begin{flushleft}
{\Large
\textbf\newline{Epidemic management and control through risk-dependent individual contact interventions} 
}
\newline
\\
Tapio Schneider\textsuperscript{1,\dag,*},
Oliver R.\ A.\ Dunbar\textsuperscript{1,\dag},
Jinlong Wu\textsuperscript{1},
Lucas B{\"o}ttcher\textsuperscript{2,3},
Dmitry Burov\textsuperscript{1},
Alfredo Garbuno-I\~{n}igo\textsuperscript{4},
Gregory L. Wagner\textsuperscript{5},
Sen Pei \textsuperscript{6},
Chiara Daraio\textsuperscript{1},
Raffaele Ferrari\textsuperscript{5},
Jeffrey Shaman\textsuperscript{6}
\\
\bigskip
\textbf{1} California Institute of Technology, Pasadena, CA, USA
\\
\textbf{2} Computational Social Science,
Frankfurt School of Finance and Management, Frankfurt a.\ M., Germany
\\
\textbf{3} Department of Computational Medicine, University of California, Los Angeles, CA, USA
\\
\textbf{4} Departamento de Estadística, Instituto Tecnológico Autónomo de México, Ciudad de México, México
\\
\textbf{5} Massachusetts Institute of Technology, Cambridge, MA, USA
\\
\textbf{6} Department of Environmental Health Sciences, Mailman School of Public Health, NY, USA

\bigskip

%
%
\dag These authors contributed equally to this work.


* tapio@caltech.edu

\end{flushleft}
\section*{Abstract}
Testing, contact tracing, and isolation (TTI) is an epidemic management and control approach that is difficult to implement at scale because it relies on manual tracing of contacts. Exposure notification apps have been developed to digitally scale up TTI by harnessing contact data obtained from mobile devices; however, exposure notification apps provide users only with limited binary information when they have been directly exposed to a known infection source. Here we demonstrate a scalable improvement to TTI and exposure notification apps that uses data assimilation (DA) on a contact network. Network DA exploits diverse sources of health data together with the proximity data from mobile devices that exposure notification apps rely upon. It provides users with continuously assessed individual risks of exposure and infection, which can form the basis for targeting individual contact interventions. Simulations of the early COVID-19 epidemic in New York City prove the concepts. In the simulations, network DA identifies up to a factor 2 more infections than contact tracing when both harness the same contact data and diagnostic test data. This remains true even when only a relatively small fraction of the population uses network DA. When a sufficiently large fraction of the population ($\gtrsim 75\%$) uses network DA and complies with individual contact interventions, targeting contact interventions with network DA reduces deaths by up to a factor 4 relative to TTI. Network DA can be implemented by expanding the computational backend of existing exposure notification apps, thus greatly enhancing their capabilities. Implemented at scale, it has the potential to precisely and effectively control future epidemics while minimizing economic disruption.

\section*{Author summary}
During the ongoing COVID-19 pandemic, exposure notification apps have been developed to scale up manual contact tracing. The apps use proximity data from mobile devices to automate notifying direct contacts of an infection source. The information they provide is limited because users receive only rare and binary alerts. Here we present network data assimilation (DA) as a new digital approach to epidemic management and control. Network DA uses the same data as exposure notification apps but uses it more effectively to provide frequently updated individual risk assessments to users. 

Network DA is based on automated learning about individuals’ risk of exposure and infectiousness from crowd-sourced health data and proximity data. The data are aggregated with models of disease transmission to produce statistical assessments of users' risks. In an extensive simulation study of the COVID-19 epidemic in New York City (NYC), we show that network DA with diagnostic testing achieves epidemic control with fewer than half the deaths that occurred during NYC’s lockdown, while isolating a far smaller fraction of the population (typically only 5–10\% of the population at any given time). 

Implemented at scale, then, network DA has the potential to effectively control epidemics while minimizing economic and social disruption.


\section*{Introduction}
Until a majority of the global population has reached immunity against continuously evolving virus variants through vaccination or infection, the ongoing COVID-19 pandemic and future epidemics will need to be fought with non-pharmaceutical interventions (NPIs) \cite{doi:10.1126/science.abd9338,haug2020ranking}. They include social distancing, mask usage, and restrictions of mass gatherings. But NPIs such as lockdowns come at catastrophic costs to individuals, economies, and societies, with disproportionate burdens carried by disadvantaged groups \cite{Konig21w,Chang21a}. Even if imposed only intermittently and regionally, lockdowns are an inefficient means of epidemic management and control: they isolate much of the population, although even at extreme epidemic peaks, only a small fraction is infectious \cite{Pei20b,Kissler20a}. If individuals who are at high risk of being infectious could be identified before they infect others, control measures could be made more efficient by targeting them to this high-risk group.  

Testing and contact tracing have been discussed and partly implemented as strategies to identify individuals who are at high risk of being infectious \cite{Hellewell20a,Ferretti20a,Kucharski20a}: testing determines who is infectious, contact tracing identifies those who may have been exposed through contact with an infectious individual, and this high-risk group is then isolated. However, controlling the COVID-19 epidemic by testing, contact tracing, and isolation (TTI) has been complicated by frequent asymptomatic and presymptomatic transmission, which support silent spread, and a short serial interval, the period between the onset of any symptoms in infector and infectee  \cite{Hellewell20a,Li20a,Peak20a,He20a}. Even in ideal scenarios, contact tracing needs to identify upward of 75\% of infections to achieve epidemic control \cite{Ferretti20a,Peak20a}. Quickly diagnosing such a large fraction of infections and manually identifying exposed individuals requires testing and a contact tracing workforce at a scale that has been challenging to realize in most countries \cite{Allen20a,Watson20a}.

To scale up the contact tracing component of TTI without a massive expansion of the workforce, exposure notification apps have been developed. They rely on proximity data from smartphones or other mobile devices to identify close contacts between users \cite{Apple-Google20a,Cencetti21a}. If an individual user is identified as being infectious, prior close contacts are notified and can then self-isolate. The exposure notification is deterministic (a user is only notified when potentially exposed), and it only uses nearest-neighbor information on the network of close contacts among users. Exposure notification apps have not seen widespread use, in part perhaps because of early implementation difficulties and privacy concerns but also because they do not provide users with information except in the rare case when they receive an exposure notification \cite{Lewis21a}. Nonetheless, where they have been used, these apps have helped prevent the spread of infections \cite{Wymant21a}.

Here we present a new and much more effective way of exploiting the same information on which exposure notification apps rely. Unlike these apps, however, this method provides users with continuously updated assessments of their individual risks. The core idea is to learn about individual risks of exposure and infectiousness by propagating crowdsourced information about infection risks over a dynamic contact network assembled from proximity data from mobile devices. Instead of the deterministic assessments of exposure notification apps, our approach exploits data from diverse sources probabilistically. Various types of information, including their uncertainties, can be harnessed. For example:
\begin{itemize}
    \item Diagnostic tests, including sensitive but slow molecular tests, less sensitive but rapid antigen tests, or pooled diagnostic tests \cite{Shental20a}.
    \item Serological tests, which indicate a reduced probability of susceptibility when antibodies specific to SARS-CoV-2 (or the causative agent of another targeted disease) are detected.
    \item Self-reported clinical symptoms, elevated body temperature readings, or other wearable sensor data, which can indicate an elevated probability of infectiousness and virus transmission  \cite{Bielecki20a,Quer21l}.
\end{itemize}
Quantification of individual risks is achieved by assimilating data into a model of virus transmission and disease progression defined on the dynamic contact network assembled from proximity data. For decision making, periodically updated individual risks of having been exposed or of being infectious take the place of the deterministic assessments in exposure notification apps. The probabilistic network approach propagates data farther along the contact network than contact tracing, consistent with models of disease progression and rates of virus transmission. It harnesses more information than contact tracing, both by being able to include diverse data sources with their uncertainties and by exploiting information inherent in the network structure itself: an individual with many contacts generally is at greater risk of having been exposed than an individual with fewer contacts \cite{Meyers05a,Pastor-Satorras15a}, and such contact rates are available from the proximity data from mobile devices. 

The network and the information it contains are dynamically updated in periodic data assimilation (DA) cycles. These cycles resemble the daily DA cycles that weather forecasting centers use operationally \cite{Bauer15a}. The quantitative information that is provided by the risk assessment platform we outline in what follows can be used in similar ways as weather forecasts: to inform personal decisions by users based on their desire to avoid risk (in the weather forecasting analogy, staying home rather than going on a mountain in the face of a likely downpour) and to inform public policy when aggregate risk measures indicate that wider mandates are necessary (analogous to evacuating a city to protect lives and avoid overwhelming public health and social infrastructures when a hurricane is likely to make landfall). 

\section*{Network data assimilation}

Our point of departure is a variant of the widely used susceptible--exposed--infectious--resistant (or recovered) (SEIR) model of epidemiology, extended through inclusion of hospitalized (H) and deceased (D) compartments to an SEIHRD model \cite{Arenas20s}. Compartmental epidemiological models have traditionally been applied on the level of aggregated individuals (e.g., the population of a city or country) \cite{Bertozzi20a}; here we follow more recent work and apply the SEIHRD on an individual level on a time-dependent contact network \cite{Pastor-Satorras15a,Kiss17a}. Each individual is represented by a node on the network; time-dependent edges between the nodes are established by close contacts between individuals, as recorded by proximity data from mobile devices. Virus transmission can occur during close contacts from infectious or hospitalized nodes to susceptible nodes, which thereupon become exposed. The probability of transmission increases with contact duration, and the transmission rate can vary from node to node and with time, for example, to reflect time-varying transmission rates resulting from virus mutations or a reduced transmission rate when masks are worn. From being exposed, nodes progress to becoming infectious, and later they may progress to requiring hospitalization, recover, or die. 

\begin{figure*}
    \begin{center}
    \includegraphics[width = \textwidth]{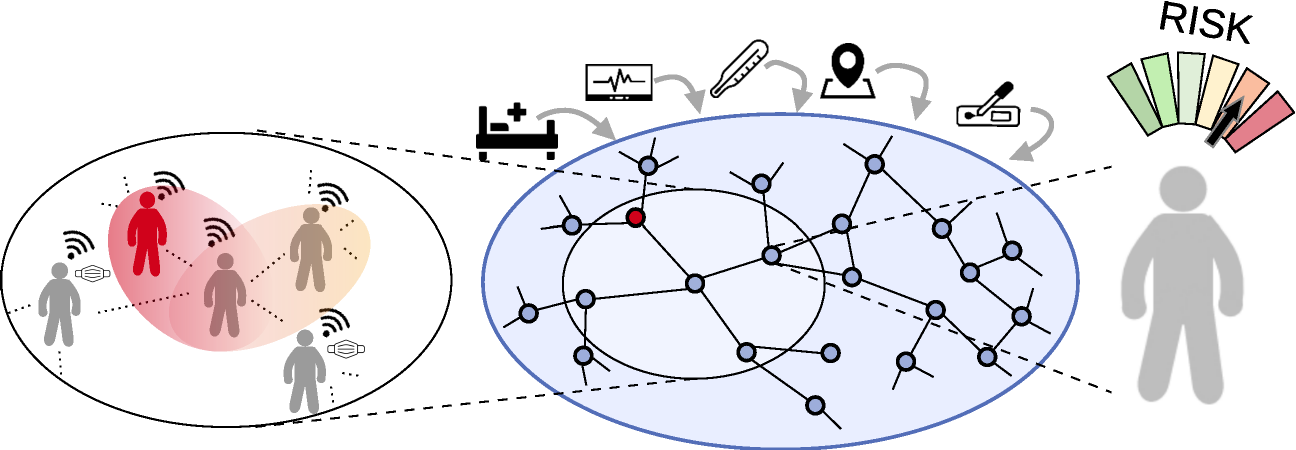}
    \end{center}
\caption{Schematic of the personalized risk assessment platform. Proximity-tracking data from mobile devices is used to assemble a contact network, in which nodes represent individuals and edges represent close contacts between individuals. An epidemiological model defined on the contact network is then fused with diverse health data, including diagnostic tests, hospitalization status, and possibly data such as body temperature readings. The model spreads risk of infectiousness from a positive individual (red) to others, taking into account knowledge about disease progression, the time and duration of contacts, and the use of personal protective equipment (PPE), among other factors. The result of the network DA is an assessment of individual risks, for example, of being infectious, which then can be used to target contact interventions.} 
  \label{f:Network_DA_Schematic}
\end{figure*}

At any time $t$, each node $i$ is in one of the six health and vital states $S_i(t)$, $E_i(t)$ etc.\ of the SEIHRD model (see Methods).  Network DA learns about the probabilities $\langle S_i(t) \rangle$, $\langle E_i(t) \rangle$, etc.\ of finding an individual node $i$ at time $t$ in each of the different states. We adopt a sequential Bayesian learning approach that propagates an ensemble of individual probabilities $\langle S_i(t) \rangle$, $\langle E_i(t) \rangle$, etc.\ across the network and periodically updates them and the SEIHRD model parameters with new data \cite{Anderson01a,Shaman12a,Pei18a,Li20a}. Data falling within a DA window of length $\Delta$ (typically, $\Delta \approx 1~\mathrm{day}$) are incorporated into the model by adjusting the ensemble to minimize the misfit to the data in the window. An interval $\Delta$ later, the updating procedure is repeated (see Methods for details). Such DA cycles and the underlying algorithms are used daily in weather forecasting to estimate up to $10^9$ variables characterizing the state of the atmosphere; they easily scale to network epidemiology models with millions of nodes or more. Essentially all types of data and their error characteristics can be assimilated with this approach, even data that are less sensitive to infectiousness, such as readings of heart rates \cite{Radin20a} or body temperatures \cite{Bielecki20a,Quer21l} (Fig.~\ref{f:Network_DA_Schematic}). 

\section*{Synthetic network for proof-of-concept}

To illustrate the methods with simulated data in a computational proof-of-concept, we construct a large synthetic contact network with $N=97,942$ nodes and about 1 million connections among them. The network has typical characteristics of a human contact network. It has a time-dependent contact rate minimum at night and a maximum midday, and it has a connectivity (degree) distribution similar to human networks: there are many individuals with few connections and a few highly connected individuals who are more likely to become superspreaders \cite{Endo21a} (Figs.~\ref{f:network_degrees}, \ref{f:dynamic_contact_network}). 

The network also contains a block representing hospitals, where hospitalized patients are connected to healthcare workers, who in turn are connected to the community in the rest of the network. Transmission rates in hospitals are reduced by a factor of 10 to reflect the use of PPE, which has proven effective in making SARS-CoV-2 transmission rates in hospitals rare (Methods). The purpose of explicitly including hospitals in the network architecture is twofold: first, to illustrate how reliable data such as hospital admittance records can be incorporated in the network DA approach; second, to enable comparison of hospitalization rates in the simulated and real epidemic while mimicking the reduced transmission rate in hospitals. Realistic human contact networks contain other structures, such as households, workplaces, and schools \cite{aleta2020modelling}. Such features are not explicitly taken into account in our synthetic network architecture; rather, contact clusters arise randomly in the synthetic network. In the real world, such clusters would arise naturally in the contact network assembled from proximity data, without the need to account for them explicitly.

As surrogates for real-world health data, we use stochastic simulations of the epidemiology model for the state variables $S_i(t)$, $E_i(t)$, etc.\ on the network. We reproduce various scenarios of the early COVID-19 epidemic in New York City (NYC), beginning during March 2020. Because age is an important risk factor for COVID-19 severity, we assign ages to nodes based on the age distribution of NYC and use them to model age-dependent disease progression according to current knowledge about COVID-19 (Tables~\ref{tab:transition_rates} and \ref{tab:hospitalization_death_rates}). While we assign ages to nodes randomly, the realism of the model could be improved with age-stratified contact patterns \cite{mistry2021inferring,monod2021age}. With the resulting surrogate worlds of contact patterns and disease progression, we explore how individual risk assessment and epidemic management and control can be achieved in what-if scenarios. 

\section*{Results}
\subsection*{Lockdown and world avoided}
\begin{figure*}
    \begin{center}
     \includegraphics[width = \textwidth]{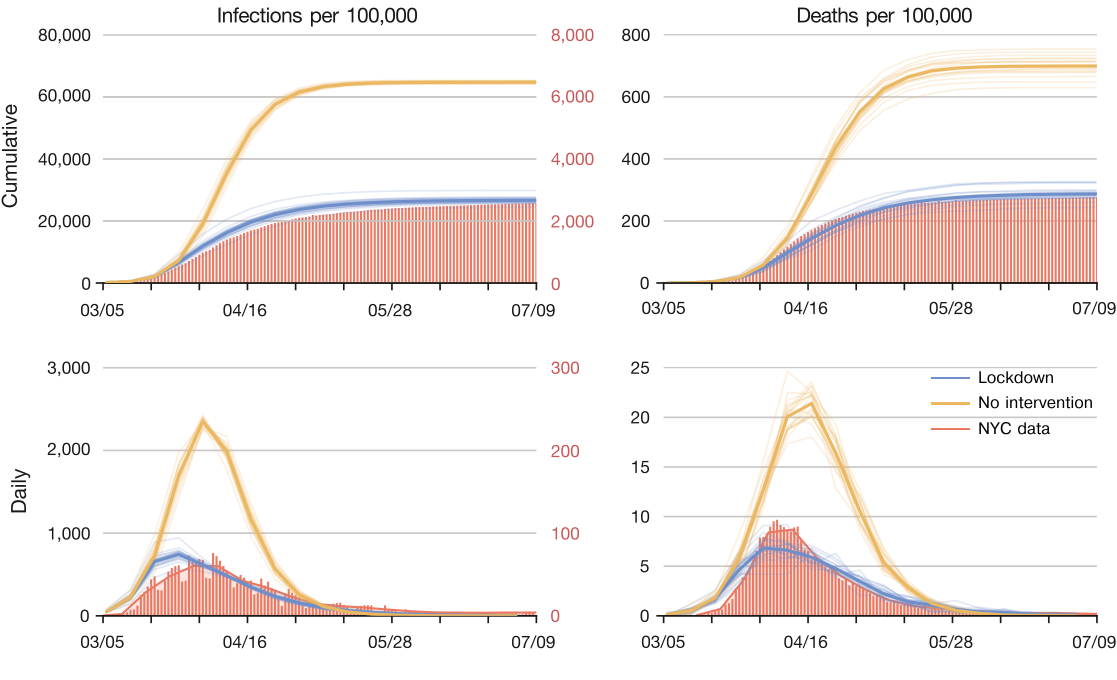}
    \end{center}
    \caption{Evolution of an outbreak in surrogate-world simulations with a lockdown (blue) and without (orange). The left column shows infection rates and the right column death rates. Upper row for cumulative counts and lower row for daily counts, smoothed with a 7-day moving average filter. Red bars represent confirmed and probable COVID-19 deaths and confirmed infection rates for New York City~\cite{nyc_covid}, with the red axis labels on the right for confirmed infection counts. Solid lines indicate the corresponding counts in the simulations, with the black axis labels on the left for infections on the network. (The axes for infections in the simulations are stretched by a factor 10 relative to the axes for confirmed NYC infections, reflecting the undercount of infections by confirmed cases \cite{Yang21e}.) The light lines show 20 simulations that only differ by random seeds, with the thicker lines indicating the ensemble mean; thus, they give an indication of sampling variability. The average contact rate for all nodes is reduced by 58\% from March 25, 2020 onward to mimic the lockdown effect (blue solid line).}
    \label{f:NYC_epidemic}
\end{figure*}

As an illustrative example, we simulate the evolution of an epidemic that, when scaled up from our network size to the NYC population of 8.3 million, resembles the early evolution of the COVID-19 epidemic in NYC in 2020 (Fig.~\ref{f:NYC_epidemic}). If the contact rate on the network remains unchanged in the simulations, the number of infections and deaths rises from early March into April, with daily deaths reaching a peak of around 21 per 100,000 population in the second half of April.

However, this world was avoided by a lockdown, which became mandatory in NYC from March 22 onward. In its wake, the number of daily new cases and deaths began to decline from mid-April onward (Fig.~\ref{f:NYC_epidemic}). We can reproduce similar behavior in the stochastic simulations by reducing the average contact rate of all nodes by 58\% starting March 25 (Fig.~\ref{f:NYC_epidemic}). The infection rates in the stochastic simulations exceed the number of confirmed cases in NYC, presumably because the latter undercount actual infections \cite{Yang21e}. However, the death rates in the stochastic simulations are close to the NYC death rate (Fig.~\ref{f:NYC_epidemic}). The hospitalization rates in the simulation also track the actual hospitalization rates closely (Fig.~\ref{fig:NYC_hospitalization}).

Thus, the simulated epidemics on the synthetic network reproduce statistics similar to the actual early epidemic in NYC, with realistic parameter choices for transmission rates and disease progression (Methods). Notwithstanding the simplifications of the network structure, this points to the qualitative adequacy of the synthetic network epidemiology model as a testbed for network DA, which makes no a priori assumptions about the structure of the network.

\subsection*{Accuracy of individual risk assessment}

To demonstrate the accuracy of individual risk assessments, we assume the network DA platform has $\tilde N \le N$ users who exchange proximity data with each other, with 25\% to 100\% of the population in the user base (i.e., $0.25 \le \tilde N/N \le 1$). In an idealization, the contact patterns of individuals within the user base are assumed to be known completely; the contact patterns of individuals outside the user base are assumed unknown. We also assume the number of external contacts of individuals in the user base to be known, for example, from proximity-sensing devices that can also detect devices of non-users. (However, we have verified that this assumption can be replaced by only assuming knowledge of the average number of external contacts, without material effects on the results; see Methods and Figs.~\ref{f:intervention_u75nbhd_avgexternal},~\ref{f:intervention_u75rand_avgexternal}.) For subsets of the $\tilde N$ users, we assimilate results of simulated rapid diagnostic tests from the corresponding nodes in the surrogate-world simulation. A fraction  $f = 1\%$, $5\%$, or $25\%$ of the user base is assumed to be tested daily, with results available on the day of test administration; that is, every user is tested on average every 100, 20, or 4 days (Methods). Testing the population of a major metropolitan center such as NYC every 100 or 20 days is achievable with current testing capacity. For more limited user bases ($\tilde N/N \le 1$), test rates of 25\% per day within the user base are locally achievable and in fact are routine, for example, on some college campuses. 

We first illustrate network DA in the worst-case scenario of the free-running synthetic epidemic, in which contact patterns do not change. DA begins on March 5. We show results for April 9, near the epidemic peak, when about 7\% of the population are infectious, and for April 30, when new infections are waning (Fig.~\ref{f:NYC_epidemic}). (In this free-running epidemic, the maximum prevalence of infectiousness is considerably higher than in the lockdown simulations, in which prevalence peaks at 1.5\%--2\%---more in line with what actually occurred during the lockdown in NYC.)

\begin{figure*}
    \begin{center}
    \includegraphics[width = \textwidth]{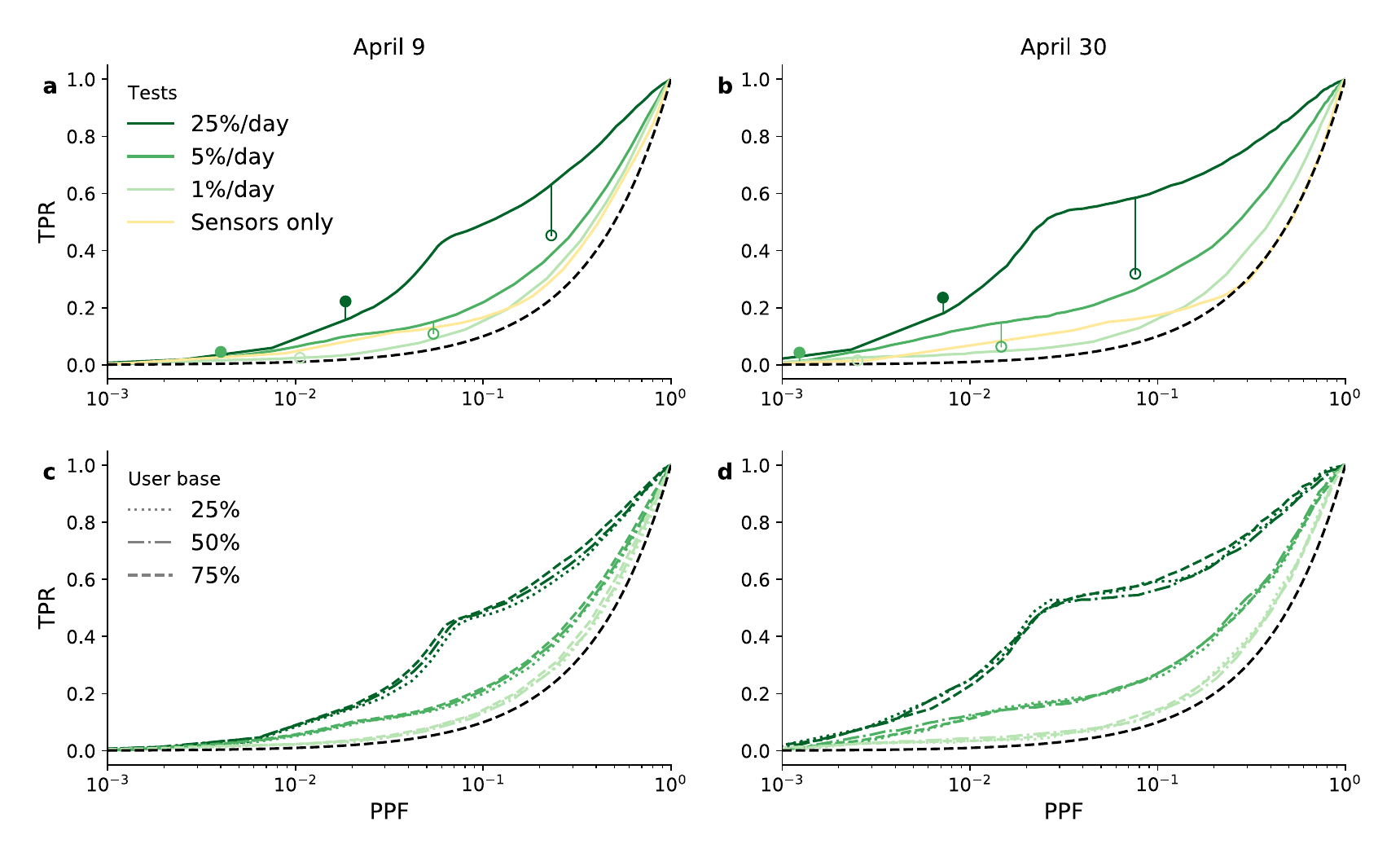}
    \end{center}
    \caption{Receiver operating characteristic (ROC) curves for classification as possibly infectious. ROC curves trace out the true positive rate (TPR) vs.\ the predicted positive fraction (PPF) as the classification threshold is varied. TPR and PPF are given relative to the user base size $\tilde N$. Green shades of the ROC curves from lighter to darker correspond to increasing diagnostic test rates $f$. Left column for April 9; right column for April 30. (a, b) For the ideal user base of $\tilde N/N= 100\%$. For comparison, the filled circles are for a test-only scenario when only users with positive diagnostic tests are classified as positive. (The 1\%/day case falls outside the plotting region; values for panel (a) are (7$\times10^{-4}$, 0.008) and for (b) are (2$\times10^{-4}$, 0.01).). The open circles are for a contact-tracing scenario in which additionally prior close contacts of users with positive diagnostic tests are classified as positive. Also shown is a sensors-only scenario in which 75\% of the user base is assumed to provide daily body temperature readings. (c, d) For user bases consisting of neighborhoods in the network covering 25\%, 50\%, and 75\% of the total population (Fig.~\ref{fig:user_base}), with the same test rates $f$ in shades of green as in (a, b). The black dashed line represents a random classifier that provides a lower bound on performance.}
    \label{f:ROC_curves}
\end{figure*}
We classify users $i$ as possibly infectious (``positive'') when the estimated probability of infectiousness $\langle I_i(t) \rangle$ exceeds a threshold $c_I$. The true positive rate (TPR)---the rate at which users who are infectious in the  stochastic surrogate-world simulation are classified as positive---naturally increases as the classification threshold $c_I$ decreases; at the same time, the positive predicted fraction (PPF)---the rate at which users overall are classified as positive, whether correctly or incorrectly---also increases because the false positive rate (FPR) increases. Receiver operating curves (ROC) trace out these competing changes in TPR and PPF (or FPR) as the classification threshold $c_I$ is varied (Fig.~\ref{f:ROC_curves}). Choosing a classification threshold $c_I$ means finding a trade-off between wanting a high TPR while keeping FPR and hence PPF low. 

In the ideal albeit unrealistic scenario when the user base encompasses the whole population ($\tilde N/N=100\%$), TPRs for April 9 are 12\%, 19\%, and 47\% for a PPF of 8\% and test rates from $f=1\%,$ 5\%, and $25\%$. Later in the epidemic, for April 30, TPRs are 13\%, 27\%, and 59\%  (Fig.~\ref{f:ROC_curves}a, b). That is, the classification results improve as the network model learns about the evolution of the epidemic. The classification results are insensitive to the user base coverage $\tilde N/N$: the accuracy of the classification does not change for user bases consisting of neighborhoods in the network covering between 25\% and 100\% of the total population, even  though the scenarios with more limited user bases only use contact information for the users, not for non-users  (Fig.~\ref{f:ROC_curves}c, d). The results are also insensitive to the user base topology (Fig.~\ref{fig:user_base}): classification performance is not substantially affected whether the user base consists of neighborhoods in the total population network (Fig.~\ref{f:ROC_curves}) or of randomly selected nodes (Fig.~\ref{f:ROC_curves_random_subnet}).

To put these results in context, compare them with the following two traditional approaches:
\begin{itemize}
\item If only users with positive diagnostic tests are classified as positive, TPRs reach 0.8\%, 4\%, and 22\% for test rates $f=1\%$, $5\%$, and $25\%$, respectively, with PPFs  0.07\%, 0.4\%, and 1.8\% on April 9 and corresponding TPRs of 1\%, 4\%, and 23\% with PPFs 0.02\%, 0.1\%, and 0.7\% on April 30 (Fig.~\ref{f:ROC_curves}a, b, solid circles). This test-only TPR is close to but slightly smaller than $f$ because the test sensitivity is less than 100\%. Classification by network DA can achieve much higher TPRs than testing alone, especially at low test rates, at the expense of increased but still modest PPFs. 
\item Contact tracing and exposure notification apps classify as positive users with positive diagnostic tests, plus their potentially exposed nearest neighbors on the network. If, following standard contact tracing protocols, individuals are classified as positive if, over the 10 days preceding the diagnosis, they had at least one contact of more than 15~minutes length with a user who had a positive diagnostic test, the so-obtained contact-tracing TPRs for April 9 are  2.4\%, 11\%, and 45\% for test rates from $f=1\%$, $5\%$ and $25\%$, with PPFs of 1\%, 5\% and 23\% (Fig.~\ref{f:ROC_curves}a, b, open circles). For April 30 the corresponding TPRs are 2\%, 6\%, and 32\% with PPFs 0.2\%, 1.5\%, and 7.6\%. Network DA exploits the same data as contact tracing and exposure notification apps but achieves substantially higher TPRs at the same PPF. For example, at the same PPF as contact tracing, network DA achieves about a 40\% higher TPR than contact tracing for April 9, and about a 100\% (factor 2) higher TPR for April 30 (Fig.~\ref{f:ROC_curves}a, b, vertical lines above open circles). That is, network DA in this synthetic example exploits the exact same data as contact tracing or exposure notification apps, but it does so much more effectively.
\end{itemize}

Network DA can also be used to assess quantitatively to what extent lower-fidelity data can improve classification. As an example, we conducted a set of experiments in which 75\% of the users were assumed to report body temperatures daily---for example, with wearable sensors \cite{Quer21l}---with infectiousness indicated by elevated temperature readings with 20\%  sensitivity \cite{Bielecki20a}. Such temperature readings improve the classification when no or few  ($f=1\%$) diagnostic tests are available; however, they do not provide a substantial benefit when $f=5\%$ of the user base or more can be tested daily (Fig.~\ref{f:ROC_curves}a, b). Nonetheless, if widely adopted, temperature sensors can provide a modest benefit when diagnostic testing capacity is low \cite{Quer21l}. 

The results show that network DA allows identification of a large fraction of infectious individuals, provided widespread testing is available. The improved identification of infectious individuals over traditional methods is insensitive to the fraction of the population covered by the user base, to the user base topology, and to stochastic variability of the epidemic. Network DA extends classification beyond the nearest network neighbors on which contact tracing and exposure notification apps focus. This gives it an advantage especially when testing capacity is limited. 

The capability of network DA to identify infectious individuals can be used to tailor individualized contact interventions for epidemic management and control. For epidemic management and control to be effective, however, it is important not only that the classification accuracy is high but also that the user base coverage is sufficiently large so that a large fraction of infectious individuals can be identified in the population, rather than just within the user base.

\subsection*{Risk-tailored contact interventions}

The individual risk assessments can be used to prompt those who are classified as possibly infectious for contact interventions. As an illustrative example of such individual contact interventions, we assume that users of the app self-isolate by reducing their contact rate with others by 91\%, to an average of 4 contacts per day, during the time when they are classified as positive and 5 days thereafter; all others in the population, whether app users or not, do not change their behavior. As a baseline for comparison, we present TTI scenarios with the same contact rate reduction but continuing over 14 days after diagnosis or identification as possibly exposed through contact with an infectious individual. For this baseline TTI scenario, an individual is classified as exposed if over the preceding 10 days, they had at least one contact lasting more than 15 minutes with an individual who had a positive diagnostic test; that is, the contact trace stage of this baseline TTI emulates techniques used in exposure notification apps, relying on the same data as those available for network DA in our synthetic examples. For a direct and fair comparison with network DA, TTI compliance is assumed to be confined to the user base. We use uniform testing regimes with test rates $f=1\%$, $5\%$, and $25\%$ within the user base. As classification threshold, we choose a fixed threshold $c_I = 1\%$, resulting in $\mathrm{TPR} \gtrsim 40\%$  and $\mathrm{PPF} \lesssim 9\%$ when contact interventions commence. Choosing the classification threshold $c_I$ adaptively, in response to current prevalence of infectiousness in the population, may further improve the results.

\begin{figure*}
    \begin{center}
    \includegraphics[width=\textwidth]{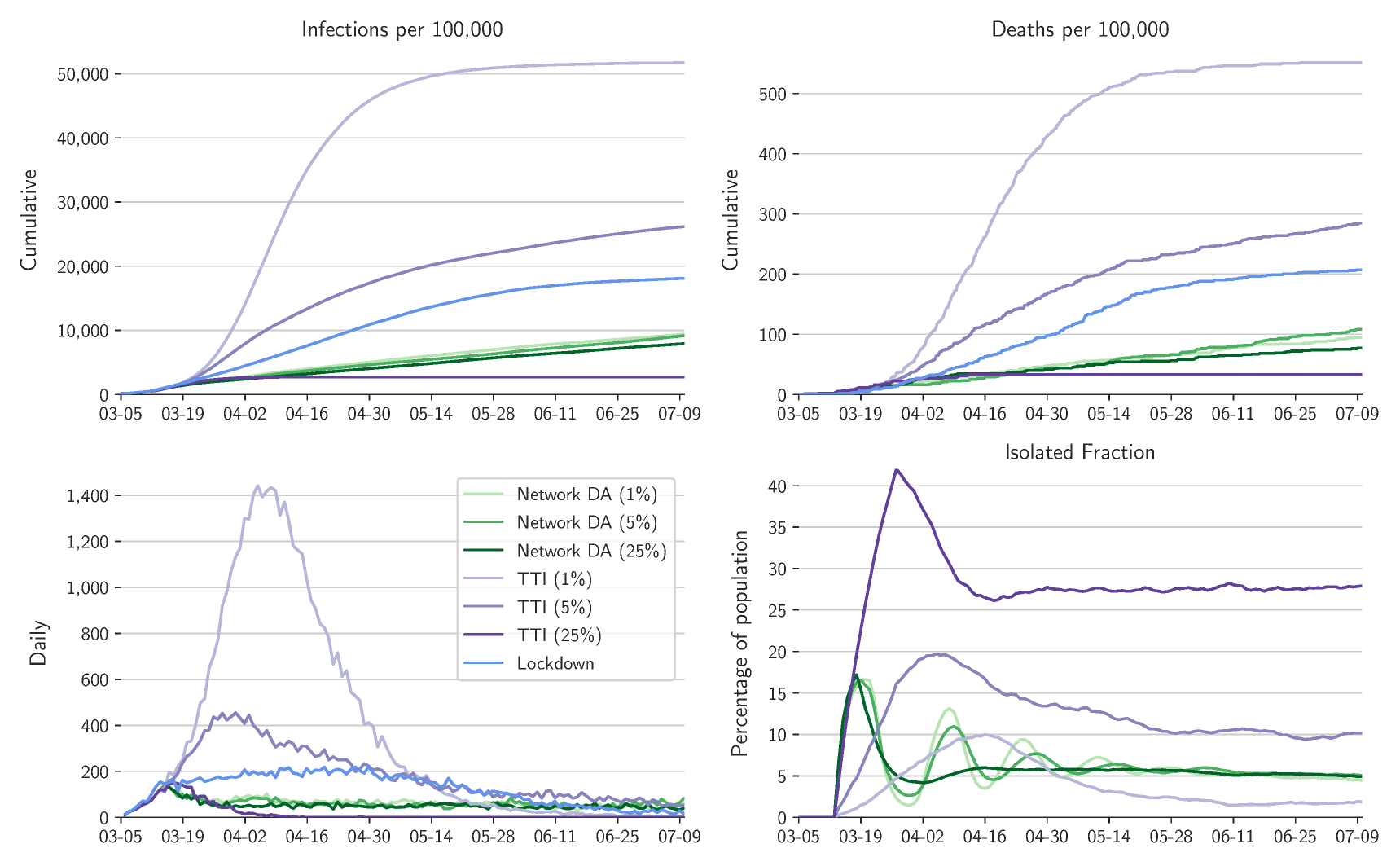}
    \end{center}
    \caption{Comparison of different contact intervention scenarios for full user base with $\tilde N/N=100\%$. Shown are the lockdown scenario (blue) from Fig.~\ref{f:NYC_epidemic}, the results of network DA and isolation of positive individuals for test rates $f=1\%$, $5\%$, and $25\%$ (greens), and the results of TTI with the same test rates as for network DA (purples).}
    \label{f:intervention_u100}
\end{figure*}

In the idealized but unrealistic case with full user base coverage ($\tilde N/N=100\%$), the epidemic is more strongly suppressed with the network DA interventions than in the lockdown scenario, with 50--70\% fewer cumulative deaths (Fig.~\ref{f:intervention_u100}). However, whereas in the lockdown scenario the entire population has reduced contacts, with network DA only a small fraction of the population self-isolates. The self-isolation fraction has an initial peak of 15--17\% for about a week and then falls quickly to 5--10\%, with damped relaxation oscillations over several weeks in the case with lower test rates ($f=5\%$);  50\% of those who isolate do so for 7 days or less, and 90\% for 14 days or less. That is, in this idealized case, risk-tailored self-isolation achieves effective epidemic control with isolation of only a small fraction of the population at a time. Network DA does not squash daily infections to zero, because the classification threshold $c_I$ was chosen as a compromise between wanting a reliable classification with a high TPR while avoiding isolation of a too large fraction of the population with a too high PPF (Fig.~$\ref{f:ROC_curves}$). For comparison, TTI with 100\% compliance does not achieve epidemic control at a test rate $f=1\%$; at a test rate $f=5\%$, cumulative deaths are 3 times higher than with network DA because TTI misses more infections than network DA. At the test rate $f=25\%$, the cumulative death rate with TTI is comparable to or lower than with network DA, but at the expense of a 2--5 times higher isolated fraction of the population.  Whereas the performance of TTI is strongly test-rate dependent, that of network DA is less sensitive to test rate, and it is always more efficient than TTI.

\begin{figure*}
    \begin{center}
    \includegraphics[width=\textwidth]{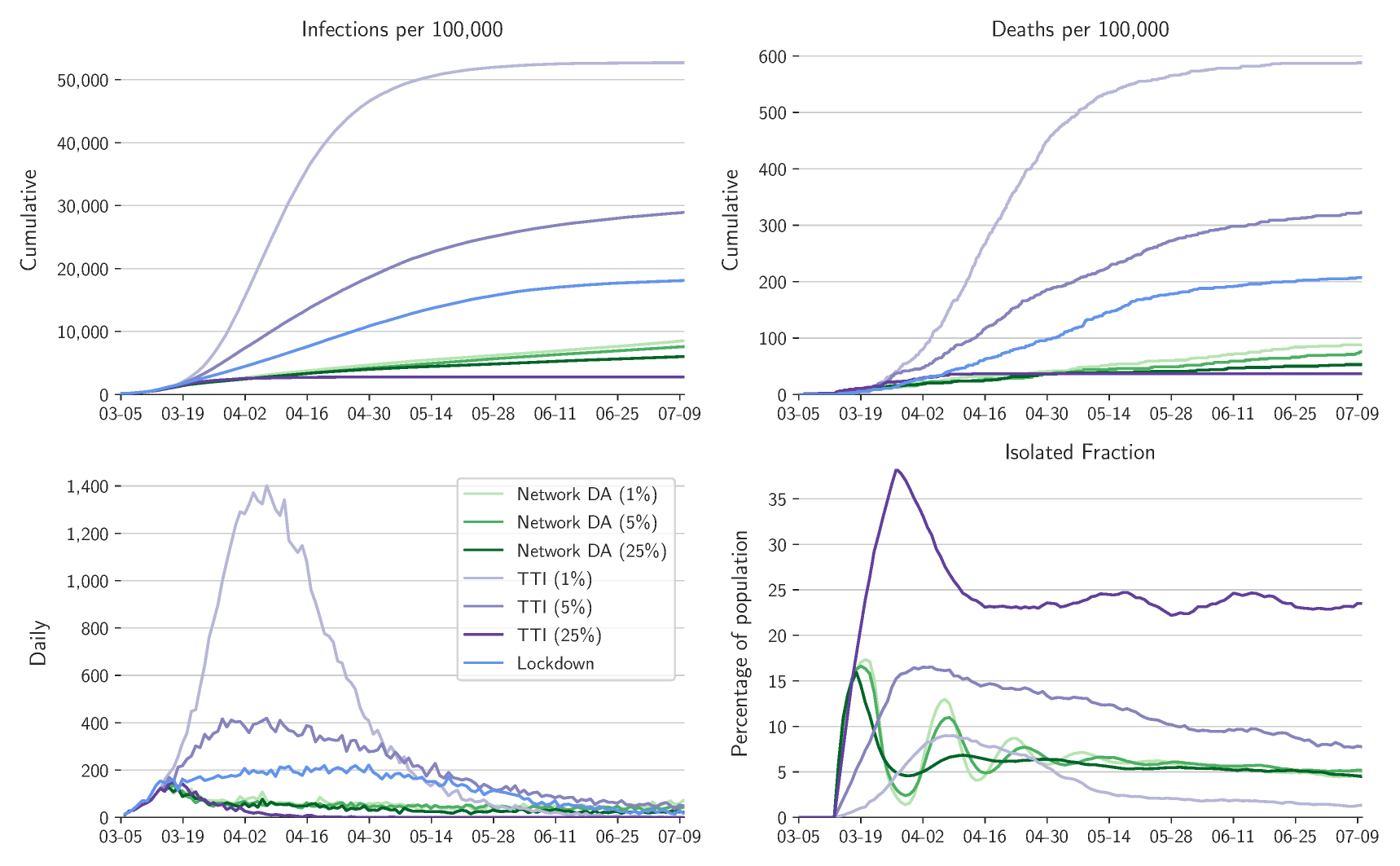}
    \end{center}
    \caption{Comparison of different contact intervention scenarios for a user base with $\tilde N/N=75\%$. Plotting conventions as in Fig.~\ref{f:intervention_u100}. TTI here is confined to the same user base as network DA, implying 75\% compliance.}
    \label{f:intervention_u75nbhd}
\end{figure*}

In the somewhat more realizable case with $\tilde N/N = 75\%$ user base coverage, we simulate a demanding scenario in which testing and contact interventions are confined to the user base; no contact information among non-users is harnessed, and non-users maintain their contact patterns without isolation. In this case, risk-tailored self-isolation still achieves epidemic control at all test rates of $f=1\%$, $5\%$, and $25\%$ within the user base (Fig.~\ref{f:intervention_u75nbhd}), and attains a cumulative death rate similar to the 100\% user base. The fraction of the population in isolation again peaks at just over 15\% initially and then drops to 5--10\%. As before, TTI with 75\% compliance and with the highest test rates ($f=25\%$) also achieves epidemic control, but with a higher isolated fraction of the population. At the test rate $f=5\%$, TTI results in an about four times higher cumulative death rate than isolation tailored by network DA, which additionally isolates fewer individuals. TTI fails to achieve epidemic control at a test rate $f=1\%$.

With a further reduced user base coverage of $\tilde N/N=50\%$, classification remains accurate, and isolation tailored by network DA can still achieve epidemic control and can remain more effective than a lockdown in preventing infections and deaths (Fig.~\ref{f:intervention_u50nbhd}). The initial fraction of the population in isolation increases to around 30\%, and then drops again to between 5--10\%. However, this means that initially, the majority of the user base (50\% of the population) is in isolation, which creates perverse incentives: it effectively puts the user base, but not others, in a lockdown. TTI with 50\% compliance fails to control the epidemic for test rates below $f=5\%$ but still achieves some control at $f=25\%$, albeit with a higher isolated population fraction than with network DA.

For the yet smaller user base coverage of $\tilde N/N=25\%$, classification remains accurate (Fig.~\ref{f:ROC_curves}); however, here the dominance of non-users within the population, who do not isolate, rules out epidemic control (Fig.~\ref{f:intervention_u25nbhd}). As with any epidemic management measure, control cannot be achieved with low compliance rates.

These results for reduced user bases are for sub-networks consisting of neighborhoods in the overall population network. Results for user bases consisting of nodes selected at random from the overall population are qualitatively similar for $\tilde N/N=75\%$, albeit with an adjusted classification threshold and a higher fraction of the population in isolation (Fig.~\ref{f:intervention_u75rand}). For $\tilde N/N=50\%$ with a random user base, network DA, while still being able to identify a large fraction of infectious individuals in the user base (Fig.~\ref{f:ROC_curves_random_subnet}), ceases to be effective for epidemic control (Fig.~\ref{f:intervention_u50rand}); similar behavior is observed in the $\tilde N/N=25\%$ case. That is, while network topology was rather unimportant for the accuracy of classification, it does play a role for the effectiveness of epidemic management and control strategies. It is possible the performance of network DA in managing the epidemic may be improved with data-adaptive classification thresholds.

\begin{figure*}
    \begin{center}
    \includegraphics[width=\textwidth]{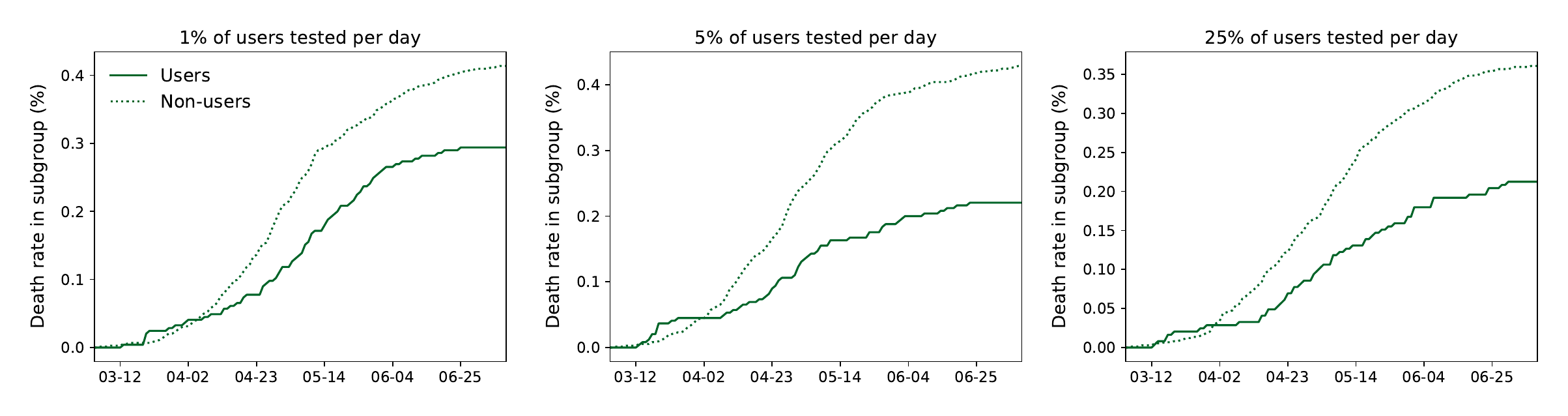}
    \end{center}
    \caption{Cumulative death rate of users vs.\ non-users for the $\tilde{N}/N=25\%$ user base consisting of neighborhoods in the overall population network. Individual contact interventions are applied within the user base from March 15 onward.}
    \label{f:user_vs_nonuser_nbhd}
\end{figure*}

In scenarios in which the user base and/or test rates are too small to achieve epidemic control, there is still a pronounced reduction in the cumulative death rate of users relative to the general non-user population (Fig.~\ref{f:user_vs_nonuser_nbhd}). For test rates $f=1\%$, $5\%$, and 25\% per day within the $\tilde{N}/N=25\%$ user base consisting of neighborhoods in the overall population network, the cumulative death rate is respectively $29\%$, $48\%$, and $42\%$  lower than the death rate among non-users. Additionally, although the contact interventions are confined to the user base, the death rate in the non-user population is still reduced by about 50\% compared with the no-intervention scenario (Fig.~\ref{f:NYC_epidemic}). For a $\tilde{N}/N=25\%$ user base consisting of nodes selected at random, the results are qualitatively similar: Death rates among users relative to non-users are reduced by $47\%$, $52\%$, and $56\%$ for respective test rates $f=1\%$, $5\%$, and $25\%$ (Fig.~\ref{f:user_vs_nonuser_rand}).

That is, risk-tailored isolation on the basis of network DA generally outperforms TTI as an epidemic management and control approach when both are presented with the same contact and test data. Even when it does not achieve epidemic control because of low compliance rates, it still offers advantages to users in terms of reduced death rates. 

\section*{Discussion}

We have demonstrated a platform concept for individual health risk assessment, which exploits the same proximity data from mobile devices that exposure notification apps rely upon but is substantially better at identifying infectious individuals. It achieves these gains by assimilating crowdsourced data from diverse sources into an epidemiological model defined on a contact network. Network DA provides informative and actionable risk assessments for individuals, even when only a modest fraction of the population uses the app necessary to obtain proximity data. The accuracy of the risk assessments is largely independent of the fraction of the population using the platform and of the user base topology; it improves with increasing diagnostic test rates, as should be expected.

When the user base is sufficiently large (covering around 75\% of the population), the platform can be used to tailor interventions that are more efficient for epidemic management and control than lockdowns or TTI. For example, with a user base covering 75\% of the population and users tested every 20 days, simulations for NYC showed that risk-tailored self-isolation achieves epidemic control with 63\% fewer deaths than during NYC's lockdown, with typically only 5--10\% of the population in isolation at any given time. This risk-tailored isolation approach is more effective at preventing infections and deaths than a TTI approach that uses the same contact and diagnostic test data. Our experiments were solely based on self-isolation among app users, without considering other public health interventions. As a result, 75\% coverage may be a conservative estimate. In reality, multiple non-pharmaceutical interventions will likely be employed simultaneously at the population level, which may reduce the user coverage required to achieve epidemic control. 

We have produced a modular codebase that allows for exploration and benchmarking of tools to manage and control epidemics in a synthetic setting. To validate and further optimize our choices of diagnostic test and intervention strategies, further analyses are required. For example, our results may be improved by the inclusion of additional information from the contact network or more data-adaptive use of the risk assessments provided by network DA. Additionally, it is possible to learn about the model parameters that appear in the network epidemiology model; we have only skimmed the surface with respect to what is possible in this regard, so far with limited success (Methods). Further investigation to delineate which model parameters are identifiable from data would be beneficial. 

The platform has a relatively low barrier to widespread implementation. It can be realized by expanding the computational backend of existing exposure notification apps. High-precision proximity data are now available through Bluetooth protocols \cite{Apple-Google20a}, and lower-precision location data from mobile devices have been exploited commercially for some time. Statistical techniques may be required to optimize the reconstruction of contact networks from such proximity data in practical implementations with imperfect knowledge of contact patterns \cite{Sim_etal20}. To be effective, the platform requires that users provide proximity data and other crowdsourced data, such as test results and reports of clinical symptoms. The more detailed data users make available, the more accurate and detailed risk profiles can be produced in return. Uptake rates of exposure notification apps have already reached up to 75\% in some urban areas, as in our simulated scenarios (e.g., more than 90\% of Singapore's population over 6 years of age \cite{TraceTogether} is using an exposure notification app). Uptake rates on a national scale so far have been more modest (e.g., a third of the UK population \cite{Wymant21a}), in part, for example, because of rural-urban digital divides but also, probably, because of the limited information provided by current exposure notification apps. However, smartphone usage rates worldwide are around 50\% and continue to grow rapidly \cite{Statista22a}; thus, widespread use of network DA in future epidemics will become technically possible. And while routine surveillance test rates in much of the world are still low, more widespread surveillance testing on the scale of major cities or regions at this point is feasible; for example, NYC currently is already testing up to roughly 2.5\% of its population daily \cite{NYCCovidTracker}. Our conclusions provide further evidence of the benefits of widespread testing, especially when that is combined with network DA to spread the test information over dynamic contact networks assembled from proximity data. 

Challenges to widespread and successful adoption of a network DA platform center around equity, compliance, and privacy questions. Smartphone use is not equitably distributed within the population, and there are disincentives (e.g., unavailability of sick leave) to comply with individual contact interventions. Conversely, classification of users as ``low risk'' may encourage risky and counterproductive behavior. It is also unknown, and we did not address, how correlations between smartphone use, compliance, and factors influencing infection risk would affect our results. An additional impediment to widespread adoption of network DA are concerns about protecting users' privacy. The network DA platform requires data to be transferred temporarily to a central computing facility for data assimilation \cite{berman2020digital}. This makes the platform more difficult to harden against malicious exploitation than exposure notification apps, which only require central data exchange when there is direct evidence of an infection \cite{Hinch20a}. Nonetheless, the data need not be stored beyond a data assimilation window that is at most a few days long. Additionally, the platform requires only anonymized proximity data but not absolute location data, and it does not rely on humans in the loop, reducing risks of malicious exploitation. There may be ways to harden the platform itself and the data exchange with users against privacy breaches \cite{Waal20a}. 

The network DA platform provides obvious benefits in managing and controlling epidemics, for example, in reducing the need for lockdowns while preventing infections and deaths, and in providing users tools to manage their personal risks. It provides a scalable alternative to manual TTI programs, and a backend that delivers more accurate and actionable information than current digital TTI and exposure notification programs developed by many governments \cite{New-York-City21s,TraceTogether}. The effectiveness of such programs has been modelled \cite{Hellewell20a,Ferretti20a,Kucharski20a}, but their impact in practice is only beginning to be elucidated \cite{Wymant21a}. Given that many TTI programs are voluntary, and documentation of contacts in manual programs is subjective, it will be important to compare both the control and cost effectiveness of manual and digital trace programs with the more objective and automated network DA approach presented here.

In addition to its health impacts, the COVID-19 pandemic has exacted an enormous economic toll on countries throughout the world \cite{Konig21w,Chang21a}. There is a continuing need to identify approaches that precisely and effectively control epidemics while minimizing economic disruption.  With sufficient uptake and testing, the platform described here provides a means for achieving these dual aims.

\section*{Methods and Models}
\subsection*{SEIHRD model on a contact network}
We consider a population of $N$ individuals $i$ (with $i=1,\dots, N$). At any time $t$, an individual $i$ is in exactly one of six possible health and vital states:
\begin{enumerate}
    \item Susceptible $S_i(t)$ when they can get infected with the virus;
    \item Exposed $E_i(t)$ when infected with the virus but not yet infectious;
    \item Infectious $I_i(t)$ when shedding the virus (with or without clinical symptoms) but not hospitalized; 
    \item Hospitalized $H_i(t)$ when hospitalized with active disease, in which case individuals are assumed to be shedding the virus;
    \item Resistant $R_i(t)$ when immune to the disease through either vaccination or immunity conferred by a prior infection (we assume lifelong resistance for now but can relax that assumption if evidence becomes available that immunity is temporary);
    \item Deceased $D_i(t)$.
\end{enumerate} 
We take $S_i(t)$, $E_i(t)$, $I_i(t)$, $H_i(t)$, $R_i(t)$, and $D_i(t)$ to be Bernoulli random variables that depend on time $t$ and take only the values $0$ and $1$. That is, $S_i(t) = 1$ when individual $i$ is susceptible at time $t$, and otherwise $S_i(t) = 0$ (and analogously for the other variables). Because the six SEIHRD states enumerate all health and vital states of individuals in this model, we have
\begin{equation}\label{e:prob_sum}
     S_i(t) + E_i(t) + I_i(t) + H_i(t) + R_i(t) + D_i(t) = 1\,.
\end{equation}
Therefore, there are only 5 independent states. 

In the network epidemiology model, a close contact between individuals $i$ and $j$  establishes a temporary network edge with weight $w_{ji}(t)=1$ for the duration $\tau$  of the contact; outside the contact period, $w_{ji}(t)=0$. Transmission along the temporary edges from one node to another and transitions between health and vital states within each node are modeled as independent Poisson processes
\cite{Meyers05a,Ferguson05a,Pastor-Satorras15a,Kiss17a}. Each process is characterized by a rate that may vary from node to node and may depend on external variables such as age, sex, and medical risk factors (see Fig.~\ref{fig:SEIRHD_schematic} for a schematic). 

We make the following assumptions about the transmission rate and the parameters characterizing transition rates between SEIHRD states, including prior distributions used in the network model for DA:
\begin{itemize}
    \item \emph{Transmission rate}: During the contact period between an infectious or hospitalized individual ($I_j(t) = 1$ or $H_j(t) = 1$) and a susceptible individual ($S_i(t)=1$), virus can be transmitted across the edge between nodes $j$ and $i$.
    When transmission occurs, the susceptible node $i$ becomes exposed and switches state to $E_i(t)=1$. During the contact period in which an edge is active ($w_{ji}(t)=1$), we assume the transmission rate to a susceptible node with $S_i(t)=1$ from an infectious node with $I_j(t) = 1$ is $\kappa_{ji}^I = a_{ji}(t)\beta$, and that from a hospitalized node with $H_j(t)=1$ is $\kappa_{ji}^H = a_{ji}'(t) \beta$. The parameter $\beta$ is a transmission rate across active edges, which we set to a global constant in the stochastic surrogate-world simulations and learn on a nodal basis in the model used for DA; $a_{ji}(t)$ and $a_{ji}'(t)$ are time-dependent transmission modifiers that can be adjusted to incorporate additional information that may be available, for example, user-supplied information that individual $i$ is using PPE at time $t$. In our proof-of-concept simulations, we use $a_{ji}(t) = 0.1$ within hospitals and $a_{ji}=1$ otherwise, to reflect the rarity of SARS-CoV-2 transmission in hospitals in which PPE is worn \cite{richterman2020hospital}. (In reality, however, depending on the types of PPE and adherence to hygiene protocols, the degree of transmissibility reduction may vary substantially among hospitals \cite{richterman2020hospital}.)
 A typical value for the transmission rate of respiratory viruses is around $\beta = 0.5~\mathrm{h}^{-1} = 12~\mathrm{day}^{-1}$ \cite{Salathe10a}. 
    
    Because we model transmission as a Poisson process, the probability that transmission occurs during contact increases with the duration of the contact period $\tau$, e.g., for an infectious node as \cite{Newman02a}
    \begin{equation*}
        T_{j i}(\tau) = 1 - e^{-\kappa_{ji}^I \tau}\,.
    \end{equation*}
    (This holds provided the contact period $\tau$ is short relative to the duration of infectiousness, so that the infectiousness status of a node does not change during contact.)
    
    In the model used for DA, we do not assume perfect knowledge of the transmission rate; instead, we learn a partial transmission rate $\beta_{i}$ for each node $i$, and compute transmission rates from node $j$ to node $i$ as the averages $\kappa_{ji}^I = 0.5 a_{ji}(t) (\beta_i +\beta_j)$ and $\kappa_{ji}^H = 0.5 a_{ji}'(t) (\beta_i +\beta_j)$. We assume independent normal priors for $\beta_i$ for each node, with a mean of $12~\mathrm{day}^{-1}$ and a standard deviation of  $3~\mathrm{day}^{-1}$. We truncate these distributions to $[1~\mathrm{day}^{-1},20~\mathrm{day}^{-1}]$, though in practice these bounds are rarely reached.
    
    \item \emph{Latent period}: Exposed nodes with $E_i(t)=1$ transition to being infectious with $I_i(t)=1$ at the rate $\sigma_i$, which is the inverse of the latent period: the time it takes for an exposed individual to become infectious. For COVID-19, the latent period lies between about 2 days and about 12~days~\cite{Lauer20a,Li20a,Kissler20a}. We take the latent period $\sigma_i^{-1}$ to be fixed for each node $i$ but heterogeneous across nodes. In the model used for DA, we represent it as $\sigma_i^{-1} = 1~\mathrm{day} + l_i$, where $l_i$ has a gamma prior distribution with shape parameter $k=1.35$ and scale parameter $\theta=2~\mathrm{day}$; hence, the minimum latent period is $1~\mathrm{day}$, and its prior mean value is 3.7~days ($1~\mathrm{day} + k\theta$).
    
    \item \emph{Duration of infectiousness in community}: Infectious nodes with $I_i(t)=1$ transition to resistant, hospitalized, or deceased at the rate $\gamma_i$, which is the inverse of the duration of infectiousness in the community (i.e., outside hospitals). For COVID-19, the median duration of infectiousness is around 3.5~days \cite{Li20a}, but its distribution has a long tail, for example, from individuals with serious or critical disease progression \cite{He20a}. Like $\sigma_i$, we take $\gamma_i$ to be fixed for each node $i$ but heterogeneous across nodes. In the model used for DA, we model the duration of infectiousness  as $\gamma_i^{-1} = 1~\mathrm{day} + g_i$, where $g_i$ has gamma prior distribution with shape parameter $k=1.1$ and scale parameter $\theta=2~\mathrm{days}$; hence, the minimum duration of infectiousness is $1~\mathrm{day}$, and its prior  mean value is $3.2~\mathrm{days}$ \cite{Li20a,He20a}. 
    
    \item \emph{Duration of hospitalization}: Hospitalized nodes with $H_i(t)=1$ transition to resistant or deceased at the rate $\gamma_i'$, which is the inverse of the duration of hospitalization. As before, we take $\gamma_i'$ to be fixed for each node $i$ but heterogeneous across nodes. In the model used for DA, we model the duration of hospitalization as $\gamma_i'^{-1} = 1~\mathrm{day} + g_i'$, where $g_i'$ has a gamma  prior distribution with shape parameter $k=1.0$ and scale parameter $\theta=4~\mathrm{days}$; hence, the minimum duration of hospitalization is $1~\mathrm{day}$, and its prior mean value is $5~\mathrm{days}$. 
    We assume hospitalized nodes are infectious. (If there is evidence that a hospitalized patient no longer sheds the virus, this can be taken into account by setting the transmission rate modifier $a_{ji}(t)$ from the corresponding node to zero; however, we are not considering this situation in our proof-of-concept.) 
    
    \item \emph{Hospitalization rate}: We assume a fraction $h_i$ of infectious nodes with $I_i(t)=1$ requires hospitalization after becoming infectious. More precisely, we assume that infectious nodes  transition to becoming hospitalized at the rate $h_i \gamma_i$. This implies that, over a period $\Delta t$ that is short relative to the duration of infectiousness $\gamma_i^{-1}$, the probability of transitioning from being infectious to hospitalized, relative to the total probability of leaving the infectious state, is
    \begin{equation*}
        \frac{1-e^{-h_i\gamma_i \Delta t}}{1-e^{-\gamma_i \Delta t}} \approx h_i \quad \text{for} \quad \gamma_i \Delta t \ll 1\,. 
    \end{equation*}
    We take $h_i$ to be fixed for each node $i$ but heterogeneous across nodes; it generally depends on age and other risk factors \cite{Arenas20s,Wu20a}. We model the age dependence in the stochastic surrogate-world simulations according to clinical data as described below (Table~\ref{tab:hospitalization_death_rates}), and we assume the same parameters in the model used for DA. 
    
   \item \emph{Mortality rate}: We assume a fraction $d_i$ of infectious nodes with $I_i(t)=1$ and a fraction $d_i'$  of hospitalized nodes with $H_i(t)=1$ die. More precisely, we assume infectious nodes die at the rate $d_i\gamma_i$, and hospitalized nodes die at the rate $d_i'\gamma_i'$. Both $d_i$ and $d_i'$ are fixed for each node but are heterogeneous across nodes, depending on age and other risk factors \cite{Arenas20s,Wu20a}. Both in the stochastic surrogate-world simulation and in the model used for DA, we assume the same age-dependent mortality rates (Table~\ref{tab:hospitalization_death_rates}).
   
    \item \emph{Resistance}: For now, we assume resistance to be lifelong, so that an individual who becomes resistant remains so indefinitely and does not return to being susceptible. This assumption can be relaxed by allowing transitions back to the susceptible state if resistance is not permanent.
\end{itemize}

The health and vital states and transition rates define a Markov chain for the individual-level SEIHRD states. The SEIHRD Markov chain on a contact network  can be simulated directly with kinetic Monte Carlo methods \cite{Gillespie77a}, as in previous studies \cite{Ferguson05a,Ferguson06a,Salathe10a,Liu18a}. We use kinetic Monte Carlo simulations both to benchmark a model for the SEIHRD probabilities and to provide a surrogate for the real world in our proof-of-concept simulations. 

\subsection*{Reduced master equations} 
We are principally interested in the individual SEIHRD probabilities, which are the expected values $\langle S_i(t) \rangle$,  $\langle E_i(t)\rangle$, etc.\ associated with the Bernoulli random variables for the states. That is, $\langle S_i(t) \rangle$ is the probability that individual $i$ is susceptible at time $t$. 

These probabilities could be obtained as averages over an ensemble of kinetic Monte Carlo simulations; however, it is more computationally efficient to solve reduced master equations for the probabilities directly. The equations are \cite{Sharkey08a,Pastor-Satorras15a}
\begin{subequations}\label{e:master_eq}
\begin{align}
    \langle \dot S_i \rangle &=  -\left[\zeta_i +  k_i^x \langle w_i \rangle P(t) \eta_i  \right]\langle S_i \rangle \,, \\
    \langle \dot E_i \rangle &=  \left[\zeta_i +  k_i^x  \langle w_i \rangle  P(t) \eta_i \right]\langle S_i \rangle   -  \sigma_i \langle E_i \rangle  \,, \\
    \langle \dot I_i \rangle &=   \sigma_i \langle E_i \rangle - \gamma_i \langle I_i \rangle \,, \\
    \langle \dot H_i \rangle &=   h_i\gamma_i \langle I_i \rangle - \gamma_i' \langle H_i\rangle \,, \\
    \langle \dot R_i \rangle &=   (1-h_i-d_i)\gamma_i \langle I_i \rangle + (1-d_i')\gamma'_i \langle H_i \rangle \,,  \\
    \langle \dot D_i \rangle &=   d_i\gamma_i \langle I_i \rangle + d_i' \gamma'_i \langle H_i \rangle \,,  
\end{align}
where
\begin{equation}\label{e:infectious_pressure}
    \zeta_i(t) = \frac{\sum_{j=1}^{\tilde{N}} w_{ji}(t) (\kappa_{ji}^I \langle S_i(t) I_j(t)\rangle + \kappa_{ji}^H \langle S_i(t) H_j(t)\rangle)}{\langle S_i \rangle}
\end{equation}
\end{subequations}
is the total infectious pressure on node $i$ from within the network formed by the $\tilde N$ users.  The infectious pressure represents the possibility of transmission to node $i$ from all network nodes that are at least temporarily connected  with node $i$. Additionally, we have included an exogenous infection rate $\eta_i$. This allows for infection from outside the network of $\tilde N$ users when the master equation network represents only a subset of a larger network with $N$ nodes, and so transmission can occur from unaccounted nodes. The exogenous infection rate $\eta_i$ is scaled by the number of external neighbors $k_i^x$ of node $i$ that are not part of the user network, by the probability $\langle w_i \rangle$ of an edge of node $i$ being active, and by the time-dependent prevalence of infectiousness $P(t)$, estimated from the network of $\tilde N$ users as described below in \eqref{e:prevalence}. The probability of exogenous infection then increases with the prevalence of infectiousness  $P(t)$ within the user base, which is taken as a proxy of prevalence outside the user base. In an idealization that may not be achievable in practice, we take the number of external neighbors $k_i^x$ as given from the network structure. In practice, the number of external neighbors can be estimated through use of the same proximity technologies (e.g., Bluetooth) on which exposure notification apps rely, which allow the sensing of other nearby mobile devices, even if they do not participate in the proximity sensing and exposure notification protocol. While this is unlikely to yield perfect knowledge about the number of external neighbors, it may be combined with statistical approximations \cite{Sim_etal20}. The net effect of these assumptions and approximations is that a user surrounded by other users will have no exogenous infection rate, while users with many external neighbors will have a larger exogenous infection rate.  We have confirmed in simulations that exact knowledge of the number of neighbors can be replaced by statistical knowledge; for example, replacing the node-dependent $k^x_i$ by the user-network average for all nodes (external connectivity in Table~\ref{tab:user_base}) yields similar results (Figs.~\ref{f:intervention_u75nbhd_avgexternal},~\ref{f:intervention_u75rand_avgexternal}). 

We integrate these ordinary differential equations with a Runge-Kutta-Fehlberg 4(5) scheme, with an adaptive timestep of maximum length 3~hours. The weights $w_{ji}(t)$ vary on shorter timescales. This is taken into account in the numerical integration by averaging $w_{ji}(t)$ over the length of a time step, rather than evaluating $w_{ji}(t)$ at discrete intervals.

\subsection*{Closure of reduced master equations}

The master equations (\ref{e:master_eq}) for the probabilities are not closed because they depend on the joint probabilities $\langle S_i(t) I_j(t)\rangle$ and $\langle S_i(t) H_j(t)\rangle$ in the infectious pressure (Eq.~\ref{e:infectious_pressure}). Various approaches to closing this term have been proposed \cite{Sharkey08a,Pastor-Satorras15a,Kiss17a}. Our approach is to estimate it from the ensemble used in the DA cycle, as follows.

The joint-event probability $\langle S_i(t) I_j(t)\rangle$ and the marginal probabilities $\langle S_i(t)\rangle$ and $\langle I_j(t)\rangle$ in the master equations (Eq.~\ref{e:master_eq}) are related through the ratio
\begin{equation}
  \mathcal{C}\bigl[S_i(t), I_j(t)\bigr] = \frac{\langle S_i(t) I_j(t) \rangle}{ \langle S_i(t) \rangle \langle I_j(t) \rangle },
\end{equation}
which is the rescaled joint probability of $S_i(t)$ and $I_j(t)$. We estimate the rescaled joint probability $\mathcal{C}\bigl[S_i(t), I_j(t)\bigr]$ by its ensemble analogue
\begin{equation}\label{e:second_moment}
     \mathcal{C}_M\bigr[S_i(t), I_j(t) \bigr] = \frac{ \overline{\langle S_i(t) \rangle \langle I_j(t) \rangle}}{\overline{\langle S_i(t) \rangle} ~ \overline{\langle I_j(t) \rangle}},
\end{equation}
where $\overline{(\cdot)} = M^{-1} \sum_m (\cdot)$ denotes the mean over the ensemble (with $m=1,\dots,M$ labeling ensemble members). Thus, we approximate the joint probability in the infectious pressure (Eq.~\ref{e:infectious_pressure}) as
\begin{equation}
    \langle S_i(t) I_j(t) \rangle =  \langle S_i(t) \rangle \, \langle I_j(t) \rangle \, \mathcal{C}_M\bigl[S_i(t), I_j(t)\bigr].
\end{equation}
With this empirical approximation, we obtain a closed-form expression for the second moment $\langle S_i^m(t) I_j^m(t) \rangle$ for each ensemble member $m$, which we use in the reduced master equations. The second moment $\langle S_i^m(t) H_j^m(t)\rangle$ follows analogously. If $\mathcal{C}_M\bigl[S_i(t), I_j(t)\bigr]=1$ and $\mathcal{C}_M\bigl[S_i(t), H_j(t)\bigr]=1$, this reduces to the mean-field approximation that is commonly made in epidemiological models \cite{Sharkey08a,Pastor-Satorras15a} and that often is accurate on real-world networks \cite{Gleeson12a}.  

We verified this closure against direct kinetic Monte Carlo simulations of the SEIHRD model on the synthetic network described below, in the free-running NYC simulation without lockdown (Fig.~\ref{f:NYC_epidemic}). The closure has similar performance as the mean-field approximation and adequately, albeit not perfectly, captures the stochastic network dynamics (Figs.~\ref{f:closure_dynamics_mf} and ~\ref{f:closure_dynamics_ec}). The closure correction coefficients (\eqref{e:second_moment}) concentrate close to the value of 1 (Fig.~\ref{f:closure_correction_coeffs}), which explains the similar performance to the mean-field approximation.

\subsection*{Data assimilation algorithm}

For DA, we use a version of the ensemble adjustment Kalman filter (EAKF) \cite{Anderson01a}, which has previously been used with epidemiological models \cite{Shaman12a,Pei18a,Li20a,Pei20b}. EAKF treats an ensemble of $M$ model parameters and states $\langle S^m_i(t) \rangle$, $\langle E^m_i(t) \rangle$, etc.\ ($m=1,\dots,M$) from a previous DA cycle as a prior and then linearly updates the ensemble of model parameters and states to obtain an approximate Bayesian posterior given the available data. Unlike other algorithms for computing Bayesian posteriors on networks \cite{Altarelli14a}, it makes no assumptions about the network structure, and it scales well to high-dimensional problems \cite{Anderson01a}.

To learn about parameters and the states of nodes on the network, we use a scheme based on iterating forward passes of the master equations over a time window $\Delta$, with EAKF updates between each pass; a similar scheme has been used in epidemiology models before \cite{Shaman12a,Pei18a,Li20a,Pei20b}. In this way, we effectively use EAKF as a smoother, harnessing all available data in a DA window $(t_f-\Delta,t_f)$. There are two parts to the DA procedure:
\begin{enumerate}
    \item Update stage: An EAKF update step is performed to assimilate all data available for the window $(t_f-\Delta,t_f)$, using the previous ensemble model run as prior. The mismatch between the simulated ensemble trajectories and the data is used to update the combined ensemble of parameters and states at the initial time $t_f - \Delta$.
    \item Forecast stage: The updated ensemble of states $\langle S^m_i(t_f-\Delta) \rangle$, $\langle E^m_i(t_f-\Delta) \rangle$, etc., with the updated model parameters, is integrated forward up to time $t_f$, to serve as prior for the next update cycle.
\end{enumerate}

EAKF relies on linear updates and assumes Gaussian error statistics. However, the forward equations (\ref{e:master_eq}) are nonlinear. As a result, the EAKF update does not always conserve total probability, in the sense that SEIHRD probabilities for each node will not always sum to 1. We therefore augment the state with an additional total probability conservation variable, with observation equal to the target probability sum 1. The Gaussian assumption is also at odds with probabilities in $[0,1]$. We have experimented with approaches of transforming variables to an unbounded space, leading to total probability conservation becoming highly nonlinear. We found it to be more robust to work in the original space where total probability conservation is a linear constraint. This approach does, however, violate Gaussianity assumptions about the ensemble when we reinforce the probability bounds by clipping the states $\langle S^m_i(t_f-\Delta) \rangle$, $\langle E^m_i(t_f-\Delta) \rangle$, etc.\ to $[0,1]$. 

We assume data errors to be uncorrelated, so that their error covariance matrix is diagonal (see below for how we specify error variances in the proof-of-concept simulations). Prior information on parameters and states is specified in EAKF through the initial condition of the ensemble. We draw the parameters of the ensemble from the above-specified prior distributions, and we initialize the state by seeding each ensemble member with a fraction of (possibly different) infectious nodes, the rest being susceptible. The initial fraction of infectious nodes is drawn from a beta distribution with shape parameters $\alpha = 0.0016$ and $\beta =1$ (not to be confused with the transmission rate $\beta$). The mean fraction (here, $0.16\%$) of initially infected nodes agrees with the fraction of initially infected nodes in the stochastic surrogate-world simulations. 

To account for the multi-fidelity nature of the assimilated data, we perform EAKF in multiple passes. This allows for better conditioned data covariance matrices and for different hyperparameter choices for the different types of data. We perform the following passes to assimilate data from the lowest to the highest fidelity:
\begin{itemize}
\item In a first EAKF pass, we update parameters and states at $t_f-\Delta$ using the poorest fidelity data (e.g., temperature sensor data), followed by a forecast over $(t_f - \Delta,t_f)$;
\item In a second EAKF pass, we update parameters and states at $t_f-\Delta$ using moderate-accuracy diagnostic test data, followed by another forecast over $(t_f - \Delta,t_f)$;
\item In a final EAKF pass, we update parameters and states at $t_f - \Delta$ using data about hospitalization and death status with small error variances, followed by a final forecast over $(t_f - \Delta,t_f)$.
\end{itemize} 

There are three well-established challenges that ensemble-based filters must tackle when assimilating a number of parameters/states that is large relative to the ensemble size \cite{Houtekamer16a}: overestimation of long-range covariances, underestimation of variances, and ensemble collapse.
\begin{enumerate}
\item To prevent spurious long-range covariances, we localize the effect of observations on states within a single node \cite{Anderson12t,Houtekamer16a}. That is, direct updates of a nodal state are only due to observations at that node during the DA window. This also provides large computational savings because EAKF updates may be performed sequentially node-by-node, in any order.
\item To prevent underestimation of variances by the finite-size ensemble, which can lead to discounting of data points \cite{Anderson01a}, and to ensure well-posedness of the matrix inversions, we use regularization of the ensemble covariance matrix $\Sigma$. If $\Lambda_{\min}$ and $\Lambda_{\max}$ denote the minimum and maximum eigenvalues of $\Sigma$, we replace $\Sigma$ in the EAKF algorithm with with $\Sigma + \max(\delta(\Lambda_{\max}-\Lambda_{\min}),\delta_{\min})I$. We choose $\delta = 5/M$ to assimilate diagnostic test data, and $\delta = 1/M$ to assimilate hospitalization/death status; $\delta_{\min}$ is taken to be the mean observational noise standard deviation of the update. 
\item To prevent ensemble collapse, we add a hybrid inflation to an assimilated state with a map $x \mapsto a(x-\bar x) + \bar x + N(0,b\bar x)$, where $\bar x$ is the ensemble mean state and $N(0, b\bar x)$ is Gaussian noise with mean zero and standard deviation $b\bar x$ \cite{Houtekamer16a}. We take $a = 3.0$ and $b=0.1$.  
\end{enumerate}

Because of the binary nature of the hospitalization and death data, we do not update these states directly; doing so can lead to shocks in the system dynamics. We only update the SEIR states $\langle S_i(t_f-\Delta)\rangle, \langle E_i(t_f-\Delta)\rangle, \langle I_i(t_f-\Delta)\rangle, \langle R_i(t_f-\Delta)\rangle$ at the beginning of a DA window $t_f-\Delta \le t \le t_f$ when assimilating hospitalization and death data that fall within the DA window. 

\subsection*{Synthetic network for proof-of-concept}

We generate a synthetic time-dependent contact network in two steps:
\begin{enumerate}
\item We generate a static degree-corrected stochastic block model (SBM)~\cite{karrer2011stochastic,peixoto_graph-tool_2014}, consisting of $N$ nodes in three groups. The three groups represent (a) hospitalized patients, (b) healthcare workers with contacts both within hospitals and in the community, and (c) the community of all remaining individuals (e.g., people in an urban environment). Hospital beds in group (a) are filled when infected nodes become hospitalized; we assume an infinite supply of hospital beds. Healthcare workers in group (b) make up 5\% of all nodes, and the remaining 95\% of nodes constitute group (c). 

We describe connections within groups (a) and (b) with an Erd\H{o}s--R\'{e}nyi model and use mean degrees of 5 in group (a) and 10 in group (b), based on a social-contact analyses in a hospital setting \cite{duval2018measuring}. Hospitalized patients in group (a) can interact only with each other and with the healthcare workers in group (b). To model the interactions between groups (a) and (b), we set the corresponding mean degrees per node to 5 for edges connecting the groups. 
We parameterize the contacts among nodes in the community group (c) with a power-law degree correction. As pointed out in \cite{danon2012social}, when groups are ignored, degree distributions associated with social interactions are well-described by a negative binomial distribution, which, for example, has also been used to describe degree distributions associated with sexual-contact networks~\cite{handcock2004likelihood}. In the presence of groups, however, degree distributions of social-interaction networks have been found to exhibit a power-law tail with an exponent of about $2.5$~\cite{danon2012social}. In accordance with the results presented in \cite{danon2012social,brown2013place}, we therefore describe parts of the synthetic contact network by a stochastic block model with power-law degree correction with exponent $2.5$, mean degree $\hat k =10$, and maximum degree $100$; Figure~\ref{f:network_degrees} shows the degree distribution.  The community (c) as a group only interacts with healthcare workers (b), and we set the corresponding mean degree to 5. 

\item To model time-dependence of the network, we make the edges of the static SBM network created in the first step time-dependent by switching them on and off. That is, neighbors of all nodes remain fixed in time, but their connections are activated or deactivated with time. We account for day/night cycles in the edge weights $w_{ji}(t)$, but we ignore, e.g., weekly cycles. We generate a diurnal cycle that replicates some properties of observed time-dependent human contact networks \cite{sociopatterns}: The fraction of active edges is small at night and in the early morning hours, reaches a maximum around noon, and approaches small values again in the evening. 

To model the time-dependence of $w_{ji}(t)$, we use a birth-death process commonly used in queuing theory. The birth-death process is a Markov chain in which arrivals (edge activations) are inhomogeneous Poisson processes with a diurnally varying mean rate $A_{ji}(t)$; contact durations are exponentially distributed with a mean contact duration $\tau$ (i.e., a mean rate parameter $\mu = \tau^{-1}$). We choose a mean duration of $\tau = 2~\mathrm{min}$ (hence $\mu = 720~\mathrm{day^{-1}}$), based on high-resolution human contact data \cite{Salathe10a}. We model the mean edge activation rate as 
\begin{equation}\label{e:mean_contact_rate}
    A_{ji}(t) = \frac{1}{\hat k} \max\Bigl\{\min(\lambda_{j, \min}, \lambda_{i, \min}), ~
        \min(\lambda_{j, \max}, \lambda_{i,\max}) 
        \left[1 - \cos^{4}\left( \frac{\pi t}{1~\mathrm{day}} \right)\right]^4 \Bigr\}.
\end{equation}
Here, $t=0$ starts at midnight, and $\hat k$ is the mean degree of the network in the community group (c), so that $\hat k A_{ji}$, when averaged over edges, is an average contact rate per node. The diurnally averaged edge activation rate then is 
\begin{equation}\label{e:daily_avg_edge_rate}
\bar A_{ji} = \frac{1}{1~\mathrm{day}} \int_0^{1~\mathrm{day}} A_{ji}(t) \,\mathrm{d}t.
\end{equation}

For the minimum and maximum contact rates per node, $\lambda_{i, \min}$ and $\lambda_{i, \max}$, we choose the default values $\lambda_{\min} = 4~\mathrm{day}^{-1}$ and $\lambda_{\max} = 84~\mathrm{day}^{-1}$. If the default contact rates apply for all nodes, this gives for the community group (c) a mean contact rate per node of 
\begin{equation}\label{e:daily_avg_contact_rate}
\hat k \bar A_{ji} \approx 37.7~\mathrm{day}^{-1};
\end{equation}
this is about a factor 3--4 larger than typical human contact rates as assessed by self-reports \cite{Mossong08a}, consistent with the fact that we also take fleeting contacts into account that would likely not be self-reported. The minimum and maximum contact rates $\lambda_{i, \min}$ and $\lambda_{i, \max}$ for a node $i$ are the principal parameters we vary to explore the effect of contact interventions. Reducing $\lambda_{i, \min}$ and $\lambda_{i, \max}$ for a node reduces the fraction of time edges connecting to node $i$ are active. The contact rate and total contact duration over the network for five simulated days are displayed in Figure~\ref{f:dynamic_contact_network}.

The time dependencies of all edges $w_{ji}(t)$ are treated as independent. We update the time-dependence of each edge at midnight every simulated day, running independent birth-death processes with parameters $A_{ji}(t)$ and $\mu$ for the next day. 
\end{enumerate}

If a node becomes hospitalized, it is deactivated at its previous location in the network and transferred to the hospital group (a). Hospitalized nodes are assumed to be infectious. (This assumption may later be relaxed to model the situation that a patient is no longer infectious but may still be hospitalized with ongoing disease.) 

Different choices of network architecture are, of course, possible and justifiable. The network merely serves to generate simulated data for our proof-of-concept, and the algorithms we demonstrate adapt to any network architecture, which in practice would be provided by proximity data. We do not expect our results to depend sensitively on our choice of network architecture.  

\subsection*{Selection of subnetwork for user base} To select a subnetwork for a user base with $\tilde N < N$ users, we construct subgraphs with two different topologies. First, a neighbor-based subgraph is constructed from a randomly selected seed user, by adding all neighbors of this user to the subgraph, then in a greedy fashion adding all neighbors of each member of this new subgraph, and so on, until a desired user population is reached. Second, a random subgraph is  constructed by randomly choosing nodes from the full network. Figure~\ref{fig:user_base} illustrates the different user base topologies, and Table~\ref{tab:user_base} summarizes their characteristics. \begin{table}
    \begin{center}
    \begin{tabular}{c c c c}
    Type & Population & Interior & Exterior connectivity  \\ \hline\hline
    75\% neighbor & 73,456 & 21,499 & 1.9 \\ 
    50\% neighbor & 48,971 & 6,301  & 5.2 \\
    25\% neighbor & 22,381 & 2,107   & 10.0 \\ \hline
    75\% random   & 73,353 & 7,061  & 3.1 \\ 
    50\% random   & 48,371 & 550    & 6.3 \\ 
    25\% random   & 24,482 & 33     & 9.3 \\
    \end{tabular}
    \end{center}
    \caption{\normalfont Details of the different user bases. The percentage represents approximately $\tilde N / N$, for the user population $\tilde N$. The interior defines how many users are completely surrounded by other users. The exterior connectivity gives the average number of exterior nodes connected to a node inside the user base.}
    \label{tab:user_base}
\end{table}

From Table~\ref{tab:user_base}, we see the topology of the user base affects the average number of external neighbors. To mitigate this effect, we take into account that users can be infected by neighbors external to the user base, through the additional infectious pressure terms in the equations for $\langle \dot S_i \rangle$ and  $\langle \dot E_i \rangle$ in Eq.~\eqref{e:master_eq}. Such neighbors are still detectable by proximity technologies (e.g., Bluetooth), but because they are not users of the network DA protocol, we do not know their current state.

\subsection*{Surrogate world simulation} To generate surrogate worlds with which to test the DA algorithm and interventions, we simulate epidemics on the synthetic network stochastically with kinetic Monte Carlo methods \cite{Gillespie77a}. For these stochastic simulations (but not for the model used for DA), we choose mean transmission and transition rates between SEIHRD states that are homogeneous across nodes, except for hospitalization and mortality rates that depend on age. The mean rates we use are based on current knowledge about COVID-19 (Table~\ref{tab:transition_rates}). We simulate 20 epidemics that only differ in their random seed. They are initialized on March 5, 2020, with 0.16\% of nodes randomly assigned to be infectious.
\begin{table}
    \begin{center}
    \begin{tabular}{cc}
    \multicolumn{1}{c}{Parameter} &  \multicolumn{1}{c}{Value} \\
    \hline\hline
        $\beta$, $\kappa^I$ &  $12~\mathrm{day}^{-1}$ \\
        $\kappa^H$       &  $0.1\beta$ \\
        $\sigma$         &  $(3.7~\mathrm{day})^{-1}$  \\
        $\gamma$         &  $(3.2~\mathrm{day})^{-1}$  \\
        $\gamma'$        &  $(5~\mathrm{day})^{-1}$  \\
        $\lambda_{\max}$ &  $84~\mathrm{day}^{-1}$  \\
        $\lambda_{\min}$ &  $4~\mathrm{day}^{-1}$  \\
    \end{tabular}
    \end{center}
    \caption{\normalfont Mean transmission and transition rates and maximum/minimum contact rates for the surrogate-world simulations with the stochastic SEIHRD model (Fig.~\ref{fig:SEIRHD_schematic}). The mean rates are taken to be the same for all nodes; hence, the nodal indices are suppressed. The latent period $\sigma^{-1}$ and duration of infectiousness in the community $\gamma^{-1}$ are approximated from those in refs.~\cite{Li20a} and \cite{He20a}; the duration of hospitalization ${\gamma'}^{-1}$ is from ref.~\cite{Richardson20a}, and the transmission rate $\beta$ is fit to be consistent with data for respiratory viruses \cite{Salathe10a} and to roughly reproduce NYC data.}
    \label{tab:transition_rates}
\end{table}

The dependencies of the hospitalization rate $h_i = h(a)$ and mortality rates $d_i = d(a)$ and $d'_i = d'(a)$ on age $a$ are estimates based on recent data (Table~\ref{tab:hospitalization_death_rates}). 
\begin{table}
    \begin{center}
    \begin{tabular}{r@{\hspace{2.5em}}rrrr}
    \multicolumn{1}{c}{Age group $a$ (yrs)} & \multicolumn{1}{c}{$f(a)$} & \multicolumn{1}{c}{$h(a)$} &  \multicolumn{1}{c}{$d(a)$} & \multicolumn{1}{c}{$d'(a)$}  \\
    \hline\hline
          0--17   & 20.7\%  & 0.2\%     &  0.0001\%     & 1.9\%  \\
          18--44  & 40.0\%  & 1.0\%     &  0.001\%      & 7.3\%  \\
          45--64  & 24.5\%  & 4.0\%     &  0.1\%        & 19.3\% \\
          65--74  & 8.3\%   & 7.6\%     &  0.7\%          & 32.7\% \\
          $\geq 75$ & 6.5\% & 16.0\%    &  1.5\%          & 51.2\% \\
    \end{tabular}
    \end{center}
    \caption{\normalfont Age-dependent mean hospitalization and mortality rates in the surrogate-world simulation. The share $f(a)$ of the population in each age group $a$ is taken from U.S.\ Census data \cite{nyc_age}. The age-dependent death rate in hospitals $d'$ is obtained from cumulative hospitalization and death rates in NYC by June 1, 2020 \cite{nyc_covid}, under the assumption that $90\%$ of deaths occurred in hospitals. Age-dependent hospitalization rates $h(a)$ and mortality rates $d(a)$ in the community (outside hospitals) are difficult to obtain directly from NYC data because of an age-dependent undercount of infections \cite{Yang21e}. We choose  hospitalization rates $h(a)$ that approximate data from France \cite{Salje20a}, adjusting the rates so that the overall hospitalization rate is $\sum_a f(a) h(a) \approx 3.1\%$, which is NYC's overall hospitalization rate if one assumes a cumulative COVID-19 incidence rate of 23\% \cite{Rosenberg20a}, together with NYC's actual hospitalization count (52,333 on June 1, 2020) and population (8.34 million) \cite{nyc_covid}. Similarly, the mortality rate in the community $d(a)$ is chosen such that the overall infection fatality rate is $\sum_a f(a)\bigl[d(a) + h(a)d'(a)\bigr] \approx 1.1\%$, which is NYC's overall infection fatality rate if one considers the same cumulative incidence of 23\% and the confirmed and probable cumulative death count from COVID-19 (21,607 by June 1, 2020).}
    \label{tab:hospitalization_death_rates}
\end{table}
To model age-dependent disease progression, we randomly assign ages to network nodes in the community group (c) according to the age distribution for NYC, as given by U.S. Census data~\cite{nyc_age}. Additional factors we neglect in our synthetic examples, such as age-dependent contact patterns, are likely small perturbing factors for the risk assessment results we show. We assign ages to nodes in the healthcare worker group (b) according to the age distribution among working-age adults (21--65 years old). Initially, there are no hospitalized nodes (i.e., group (a) is empty).

The network has 97,942 nodes (with the difference to 100,000 arising from stochastic effects in the generative algorithm).  We choose the global mean transmission rate $\beta$ so that our simulations are qualitatively aligned with the evolution of the COVID-19 epidemic in NYC~\cite{nyc_covid}. We find that a global value of $\beta=12~\mathrm{day^{-1}}$ can qualitatively reproduce the observed rate of infections and can quantitatively reproduce the rate of hospitalizations and deaths during the initial phases of the epidemic in NYC.

\subsection*{Synthetic data}

We sample synthetic data from the stochastic surrogate-world simulation on the network with $N$ nodes and assimilate data for a possible subset of $\tilde N \le N$ users in the reduced master equation model. We consider the following data and error rates: 
\begin{itemize}
    \item A positive virus test for node $i$ is taken to imply 
    \begin{equation}
        \langle  I_i(t) \rangle = \mathrm{PPV}
    \end{equation} 
    at the time the test sample is taken. 
    The positive predictive value (PPV) is calculated as 
    \begin{equation}\label{e:PPV}
    \mathrm{PPV} = \frac{\mathrm{sensitivity} \times P(t)}{\mathrm{sensitivity} \times P(t) + (1-\mathrm{specificity}) \times (1-P(t))},
    \end{equation}
    where we take the sensitivity of the test to be 80\% and the specificity to be 99\%, which we use as an approximation of the currently imprecisely known actual sensitivities and specificities \cite{Sethuraman20a,Wang20a}. As an estimate of the infectiousness prevalence $P(t)$ in the population, we use the average of the infectiousness probabilities both over the network of size $\tilde N$ and over the ensemble of size $M$,  
    \begin{equation}\label{e:prevalence}
    P(t) = \max \left( \frac{1}{\tilde N M} \sum_{m=1}^M \sum_{i=1}^{\tilde N} \langle I_i^m(t) \rangle, \frac{1}{\tilde{N}} \right).
    \end{equation}
    The cutoff of ${\tilde N}^{-1}$ is included to guard against erroneously assuming prevalence to be zero because of subsampling on the reduced network. For the DA, we assume an error rate of $1-\mathrm{PPV}$ for a positive test result. 
    \item A negative virus test for node $i$ is similarly taken to imply
     \begin{equation}
        \langle  I_i(t) \rangle = \mathrm{FOR}
    \end{equation} 
   at the time the test sample is taken. The false omission rate (FOR) is calculated as 
   \[
      \mathrm{FOR} = \frac{(1-\mathrm{sensitivity}) \times P(t)}{(1-\mathrm{sensitivity}) \times P(t) + \mathrm{specificity} \times (1-P(t))},
    \]
   with the same sensitivity, specificity, and prevalence as for a positive virus test. For the DA, we assume an error rate equal to $\mathrm{FOR}$ for a negative test result. 
    \item To assimilate low-fidelity data such as those from temperature sensors, we assume $\langle  I_i(t) \rangle = \mathrm{PPV}$ as for a positive virus test when they indicate infectiousness (e.g., when a temperature reading is elevated). However, we use a sensitivity of 20\% and specificity of 98\% to reflect the lack of sensitivity of temperature sensors in detecting COVID-19 infection \cite{Bielecki20a}. For the DA, we assume an error rate equal to $1-\mathrm{PPV}$, analogous to a positive virus test.
    \item Data about hospitalization with COVID-19 imply that $H_i(t) = 1$ for the duration of hospitalization. We assume the hospitalization status of all users to be known with certainty, that is, we assimilate the hospitalization status $H_i(t)=0$ or $H_i(t)=1$ for all users; however, we only update the SEIR probabilities at the beginning of a DA window $t_f - \Delta \le t \le t_f$ with hospitalization data.
    \item Death implies $D_i(t) = 1$. We assume the vital status of all users to be known with certainty, that is, we assimilate $D_i(t)=0$ or $D_i(t)=1$ for all users; as for hospitalization, however, we only update the SEIR probabilities at the beginning of a DA window $t_f - \Delta \le t \le t_f$ with death data.
    \item For completeness, we state that a positive serological test for SARS-CoV-2 for node $i$ would be taken to imply 
    \[
    \langle  R_i(t) \rangle = \mathrm{PPV},
    \]
    with the positive predictive value calculated from \eqref{e:PPV}. Typical values for sensitivity would be 90\% and for specificity 95\% \cite{fda_serological}, and the prevalence of resistance can be estimated from the resistance probabilities on the reduced master equation network. We would again assume an error rate equal to $1-\mathrm{PPV}$. However, we did not assimilate simulated serological tests in our proof-of-concept because currently achievable serological test rates are low.
\end{itemize}

To model data errors, we randomly corrupt the synthetic data sampled from the surrogate world network with the false positive and false negative rates implied by the sensitivity (false negative rate = $1 - \mathrm{sensitivity}$) and specificity (false positive rate = $1 - \mathrm{specificity}$).

\paragraph*{Testing strategy}

We use a simple testing strategy that randomly tests a given budget of nodes once per day. Our framework provides a testbed for different strategies. We found random testing consistently outperformed three other simple strategies: (i) concentrating the test budget on near-neighbors of positively tested nodes; (ii) continuous testing of a fixed subset of the population; and (iii) testing nodes with high predicted infectiousness values. We attribute this to the low prevalence of disease. However, it is possible that more effective testing strategies can be discovered that exploit estimated nodal states, the network structure, and the intervention strategy. In a real-world scenario, systematic biases in testing (e.g., testing biased toward certain workplaces or educational institutions) may also affect quantitative details of our results. 

\paragraph*{Parameter Learning}

In addition to assimilating probabilities of SEIHRD states, we can in principle learn about parameters in the reduced master equation model (\ref{e:master_eq}), for example:
\begin{itemize}
    \item Individual partial and time-dependent transmission rates $\beta_{i}$;
    \item Individual inverse latent periods $\sigma_i$;
    \item Individual inverse durations of infectiousness $\gamma_i$ and hospitalization $\gamma_i'$;
    \item Individual hospitalization  rates $h_i$ and mortality rates $d_i$ and $d_i'$;
    \item Exogenous infection rates $\eta_i$.
\end{itemize}
We have not fully explored the efficacy of learning about the different parameters from data. For now, we include only the partial transmission rates $\beta_i$, the inverse latent periods $\sigma_i$, and the durations of infectiousness $\gamma_i$ and hospitalization $\gamma'_i$ in the DA, all with the priors stated above. Figure~\ref{fig:param_posterior} shows the prior distributions of the parameters at the beginning of the epidemic, as well as the posterior distributions as the epidemic evolves and the network model learns about the parameters. The results show that the DA does not refine the prior estimates of the parameters. When priors were not initially centered on the true values, they remained biased during the simulation. Further investigations focusing, for example, on learning statistical averages of parameters rather than individual node-per-node parameters would be beneficial.

The hospitalization rates $h_i$ and mortality rates $d_i$ and $d_i'$ are fixed at the same values as in the stochastic surrogate-world simulation (Table~\ref{tab:hospitalization_death_rates}). We assume the exogenous infection rates $\eta_i$ to be equal to the partial transmission rates $\beta_i$, and we estimate the probability of an edge of node $i$ being active as
\begin{equation}
    \langle w_i \rangle = \frac{\bar A_{i}}{ \mu + \bar A_{i}},
\end{equation} 
where $\bar A_i$ is the diurnally averaged edge activation rate, 
\begin{equation}\label{e:node_mean_contact_rate}
 \bar A_{i} = \frac{1}{1~\mathrm{day}} \int_0^{1~\mathrm{day}} A_{i}(t) \,\mathrm{d}t, 
 \end{equation}
 with
 \begin{equation}
 A_{i}(t) = \frac{1}{\hat k} \max\left\{\lambda_{i, \min},\lambda_{i,\max}
        \left[1 - \cos^{4}\left( \frac{\pi t}{1~\mathrm{day}} \right)\right]^4 \right\}.
\end{equation}
With our parameters, this is $\bar A_i = \lambda_{i,\min}/\hat k = 0.4~\mathrm{day^{-1}}$ for isolated nodes and  $\bar A_i = 3.77~\mathrm{day^{-1}}$ otherwise. For a stationary birth-death process, this estimate for $\langle w_i \rangle$ is the stationary probability of an edge being active; it approximates the diurnally averaged probability in the case of the birth-death process with diurnally varying edge activation rates. Through this probability $\langle w_i \rangle$, the exogenous infectious pressure depends on the isolation status of a node.

\paragraph*{Classification in Network DA} 

Nodes $i$ in the community group (c) are classified as possibly infectious ($\mathcal{I}_i = 1$) or not ($\mathcal{I}_i = 0$) according to 
\begin{equation}
    \mathcal{I}_i = 
    \begin{cases} 
    1 & \text{if } \overline{\langle I_i^m \rangle} > c_I,\\
    0 & \text{otherwise}.
    \end{cases}
\end{equation}
Here, $c_I$ is a classification threshold, which can be determined from receiver operating characteristic (ROC) curves as some optimum tradeoff between wanting to achieve high true positive rates while keeping false positive rates modest. The ROC curves we use are adapted to the setting in which we are primarily interested in the fraction of users that is classified as possibly infectious (and thus may be asked to self-isolate). We plot the true positive rate (TPR, nodes with $\mathcal{I}_i =1$ for which $I_i = 1$ in the stochastic simulation) against the predicted positive fraction (PPF, fraction of nodes with $\mathcal{I}_i =1$ in the user base of size $\tilde N$). ROC curves are traced out by lowering the classification threshold $c_I$, thereby increasing both TPR and PPF.

\paragraph*{Classification in TTI}
For the TTI scenarios, we assume the dynamic contact network among users is known, as in the network DA scenarios, and we assume instantaneous tracing. When a node $i$ in the community group (c) is tested positive, it is classified as infectious; all nodes that have had at least one 15-minute contact with node $i$ within the preceding 10 days are classified as exposed. All infectious and exposed community nodes are immediately isolated. This TTI scheme mimics the methods of typical exposure notification apps; although it is idealized and overestimates TTI performance achievable in practical settings \cite{Ferretti20a,Con_etal21}, it provides a fair baseline for comparison with network DA. 

\paragraph*{Contact Interventions} 
We implement two types of intervention scenarios in our test cases. In the first, a lockdown scenario (Fig.~\ref{f:NYC_epidemic}), we set $\lambda_{i, \max}$ for all nodes in the community group (c) to $33~\mathrm{day^{-1}}$. This amounts to a reduction of the mean contact rate (\eqref{e:daily_avg_contact_rate}) in group (c) by 58\%. In the second, a time-limited isolation intervention, we reduce the contact rates of targeted high-risk nodes by setting $\lambda_{i, \max} = \lambda_{i, \min} = 4~\mathrm{day^{-1}}$; thus, these high-risk nodes are assumed to self-isolate, with only 4 contacts per day on average, corresponding to a reduction of their average contact rate by $91\%$. 

The duration of contact reduction takes three possible values. In the lockdown scenario (Fig.~\ref{f:NYC_epidemic}), all nodes have contact reduction from the inception of the lockdown until the end of the simulation. In the TTI scenario, self-isolating nodes have contact reduction for 14 days, in accordance with current CDC guidelines \cite{CDC-contact-tracing-21}, after which contact rates are reset to the original values. For the network DA scenario, self-isolating nodes have contact reduction until they are classified as negative ($\overline{\langle I_i^m \rangle}\leq c_I$) for a period of 5 consecutive days,  after which contact rates are reset to their original values. This corresponds to reinstatement of original contact rates for 50\%, 90\% and $>$99\% of isolated nodes within 7, 14 and 21 days, respectively.

\paragraph*{Acknowledgments}
We thank Tobias Bischoff, Mason Porter, and Andrew Stuart for helpful discussions. 

\paragraph*{Financial Disclosure Statement} 
This research was supported by Eric and Wendy Schmidt and Schmidt Futures (T.S., O.R.A.D., J.W., D.B.); Swiss National Science Foundation (P2EZP2\_191888), National Institutes of Health (R01HL146552), and Army Research Office (W911NF-18-1-0345) (L.B.); National Science Foundation (DMS-2027369), National Institute of Allergy and Infectious Diseases (R01AI163023) (S.P., J.S.) and the Morris-Singer Foundation (J.S.). The funders had no role in study design, data collection and analysis, decision to publish, or preparation of the manuscript. The California Institute of Technology has filed a provisional patent application covering the epidemic management and control methods described in this paper. JS and Columbia University disclose partial ownership of SK Analytics.

\paragraph*{Author Contributions}
Conceptualization: T.S., C.D.; Methodology: T.S., O.R.A.D., J.W., L.B., S.P., A.G.-I., J.S.; Software: O.R.A.D., J.W., L.B., D.B., A.G.-I., G.W. Formal analysis: O.R.A.D., J.W., L.B., T.S., S.P., J.S. Investigation: O.R.A.D., J.W., L.B., A.G.-I., D.B.; Writing--original draft preparation: T.S., J.S., L.B., O.R.A.D.; Writing, review and editing: T.S., J.S., O.R.A.D., L.B., J.W., A.G.-I., C.D., R.F.; Project Administration, Supervision: T.S.

\paragraph*{Data Availability Statement} All code written in support of this publication is publicly available at https://github.com/tapios/risk-networks.


\bibliographystyle{plos2015}
\bibliography{lit,covid_refs}

\clearpage

\section*{Supporting Figures}

\setcounter{figure}{0}
\renewcommand{\thefigure}{S\arabic{figure}}%


\begin{figure}[htb]
    \begin{center}
    \includegraphics[width = 2.5in]{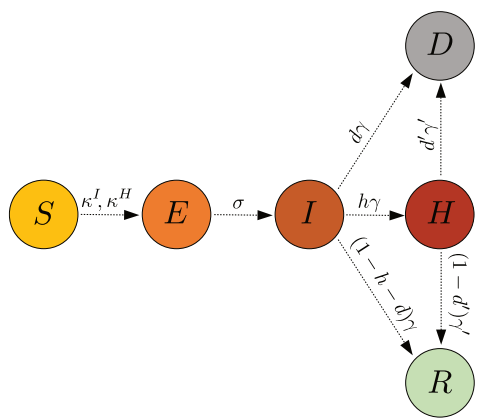}
    \end{center}
    \caption{Schematic of SEIHRD model \cite{Arenas20s}. Infected and hospitalized nodes infect susceptible nodes at rates $\kappa^I$ and $\kappa^H$, respectively. After being infected, susceptible nodes become exposed. Exposed nodes become infectious at rate $\sigma$. Infected nodes may get hospitalized at rate $h \gamma$, die at rate $d \gamma$, or become resistant at rate $(1-h-d)\gamma$. Once hospitalized, nodes either become resistant at rate $(1-d')\gamma'$ or die at rate $d' \gamma'$.}
    \label{fig:SEIRHD_schematic}
\end{figure}

\clearpage

\begin{figure*}
    \begin{center}
    \includegraphics[width=0.95\textwidth]{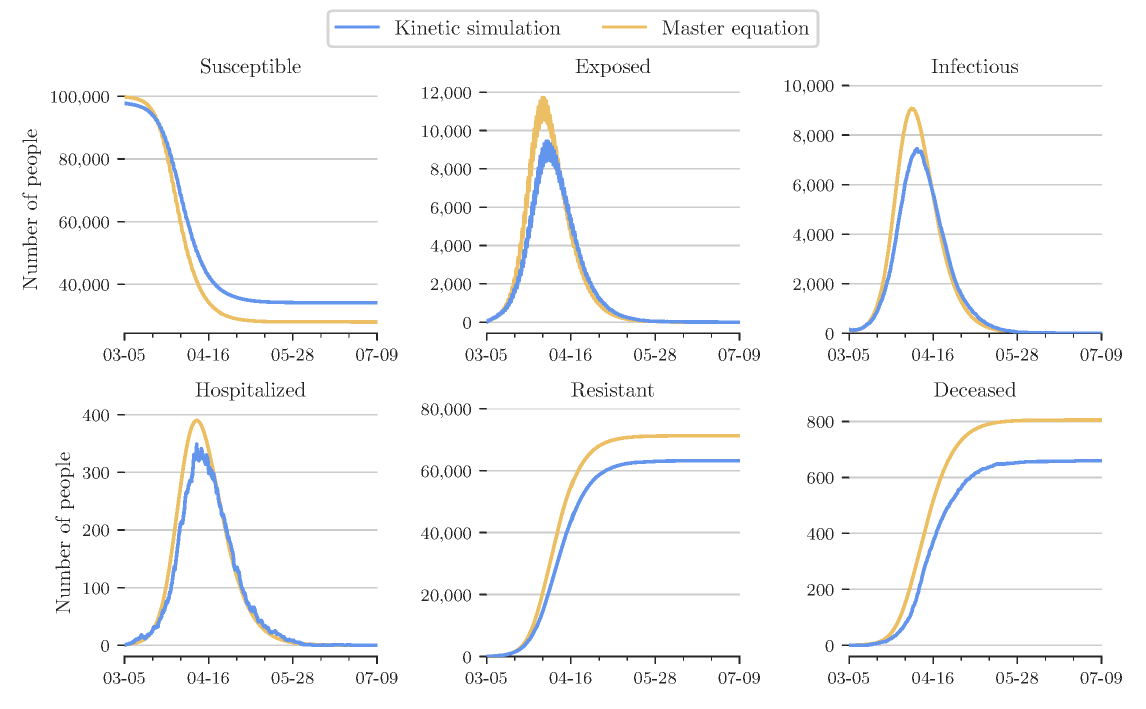}
    \end{center}
    \caption{Overall epidemic dynamics from SEIHRD model using mean-field approximation.}
    \label{f:closure_dynamics_mf}
\end{figure*}

\clearpage

\begin{figure*}
    \begin{center}
    \includegraphics[width=0.95\textwidth]{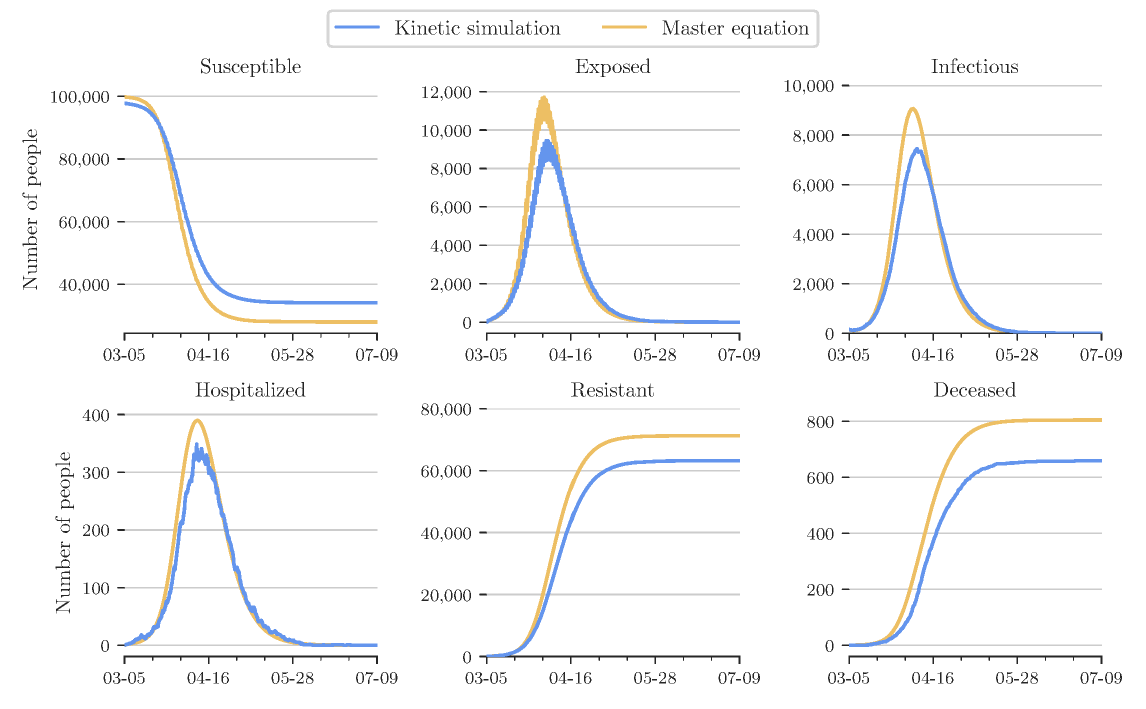}
    \end{center}
    \caption{Overall epidemic dynamics from SEIHRD model using mean-field approximation with ensemble correction.}
    \label{f:closure_dynamics_ec}
\end{figure*}

\clearpage

\begin{figure*}
    \begin{center}
    \includegraphics[width=\textwidth]{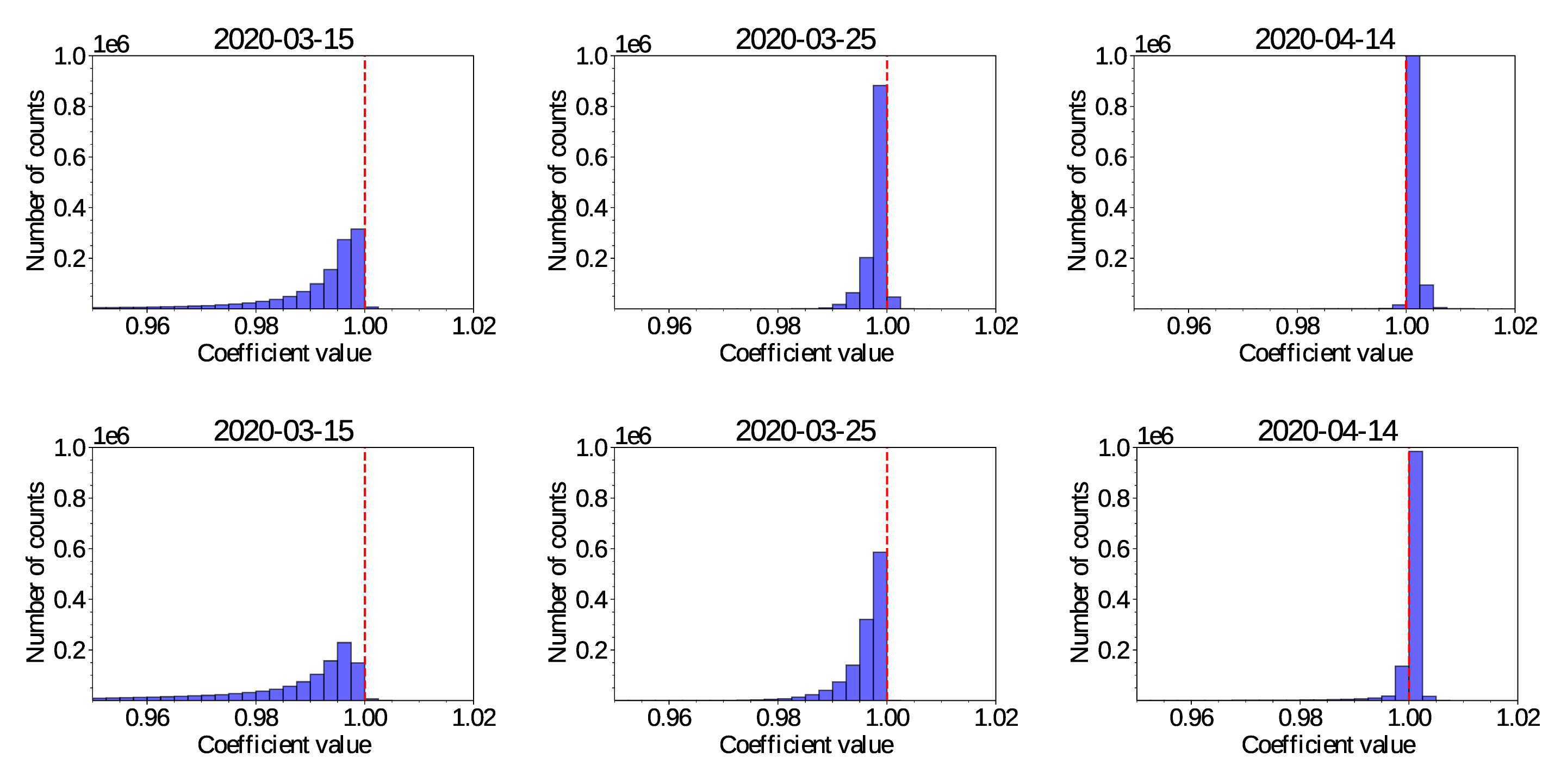}
    \end{center}
    \caption{Histograms of correction coefficients  (top row) $\mathcal{C}_M\bigl[S_i(t), I_j(t)\bigr]$ and (bottom row) $\mathcal{C}_M\bigl[S_i(t), H_j(t)\bigr]$ at different times during the simulated epidemic.}
    \label{f:closure_correction_coeffs}
\end{figure*}

\clearpage

\begin{figure}
    \begin{center}
    \includegraphics{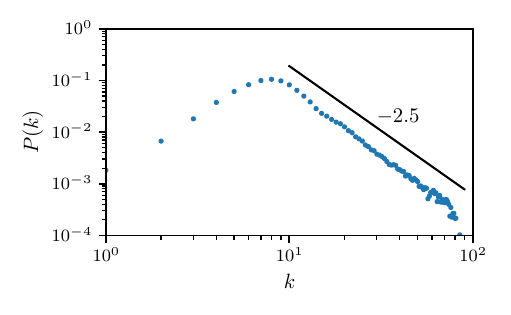}
    \end{center}
    \caption{Distribution of degrees $k$ in synthetic contact network with 97,942 nodes.}
    \label{f:network_degrees}
\end{figure}

\clearpage 

\begin{figure}
    \begin{center}
    \includegraphics[width=0.8\textwidth]{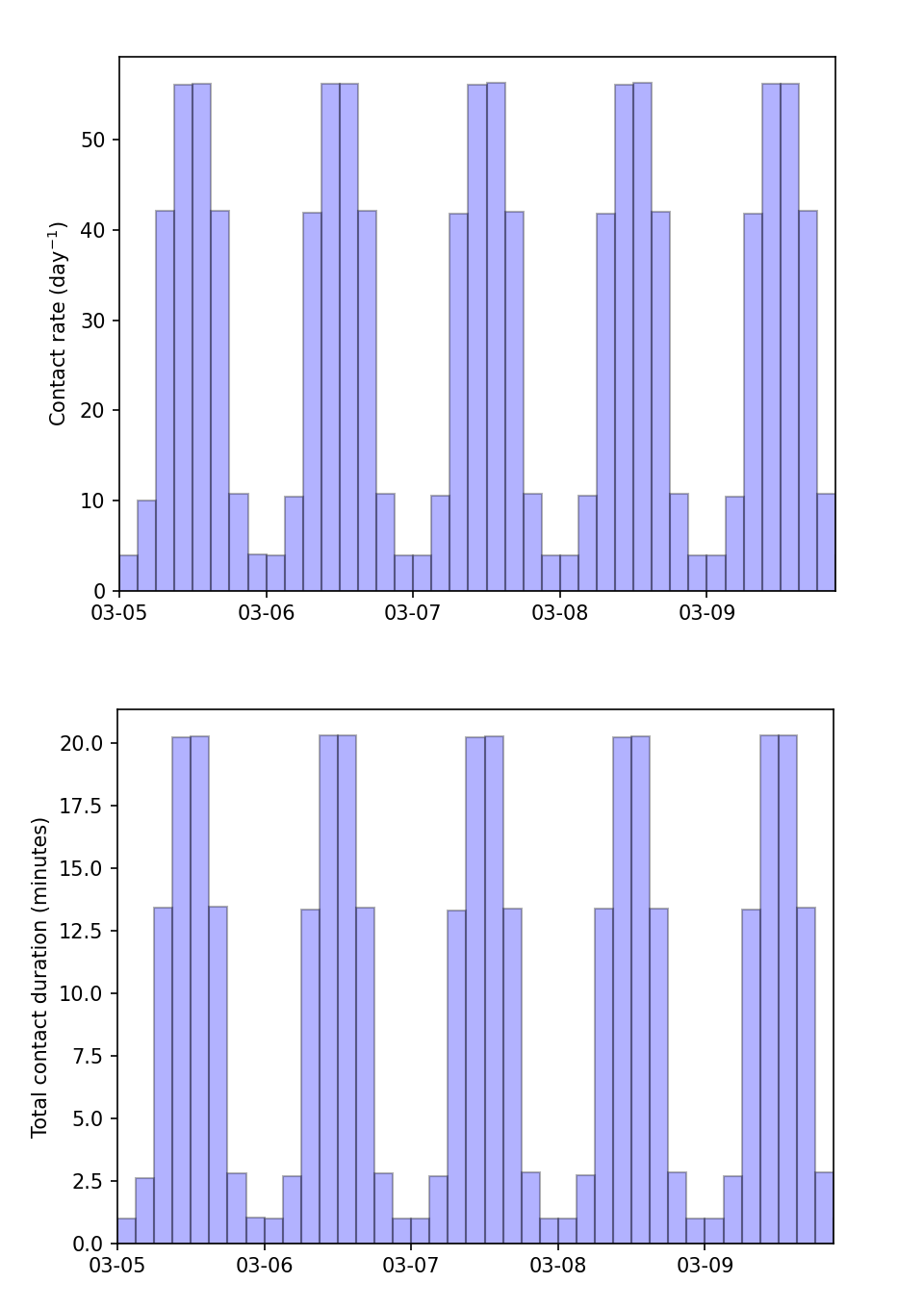}
    \end{center}
    \caption{Dynamic contact network behavior in the first five  simulated days, batched into 3-hour windows (starting at midnight). Displayed are the ensemble-averaged and node-averaged contact rate and total contact duration.}
    \label{f:dynamic_contact_network}
\end{figure}

\clearpage

\begin{figure*}
    \begin{center}
    \includegraphics[width = .8\textwidth]{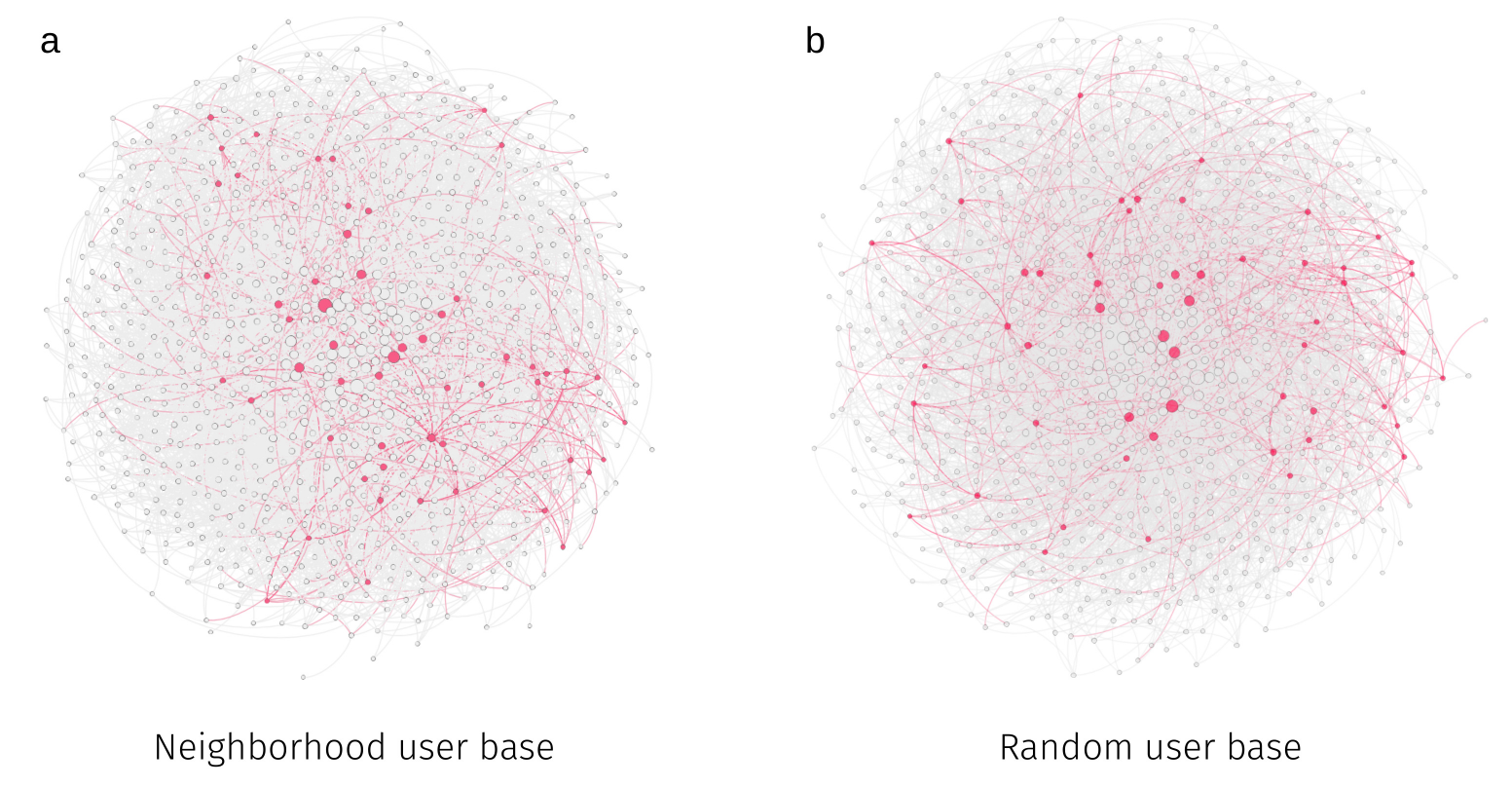}
    \end{center}
    \caption{Illustration of different user base topologies. (a) Neighbor-based user base, constructed by iteratively adding neighborhoods. (b) Random subnetwork of users. Red nodes and edges are part of a user base, grey nodes and edges of the overall population. The shown networks have 982 nodes and 5,916 edges. Both user bases contain 5\% of all nodes.}
    \label{fig:user_base}
\end{figure*}

\clearpage

\begin{figure*}
    \begin{center}
    \includegraphics[width = 0.95\textwidth]{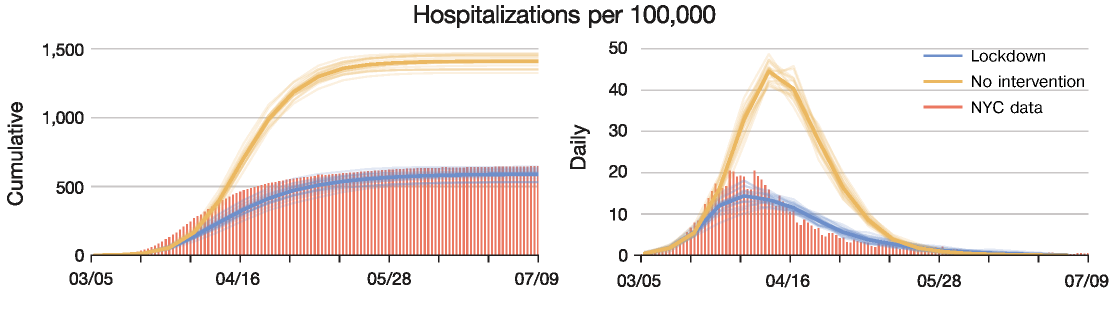}
    \end{center}
    \caption{Hospitalization rates in surrogate-world simulation with a lockdown (blue) and without (orange). The left panel shows cumulative hospitalizations and the right panel daily hospitalizations per 100,000 population for the same simulations as those in Fig.~\ref{f:NYC_epidemic}. Red bars represent COVID-19-related hospitalization rates for New York City~\cite{nyc_covid}. As in Fig.~\ref{f:NYC_epidemic}, the simulation data are smoothed with a 7-day moving average filter.}
    \label{fig:NYC_hospitalization}
\end{figure*}

\clearpage

\begin{figure*}
    \begin{center}
    \includegraphics[width=0.95\textwidth]{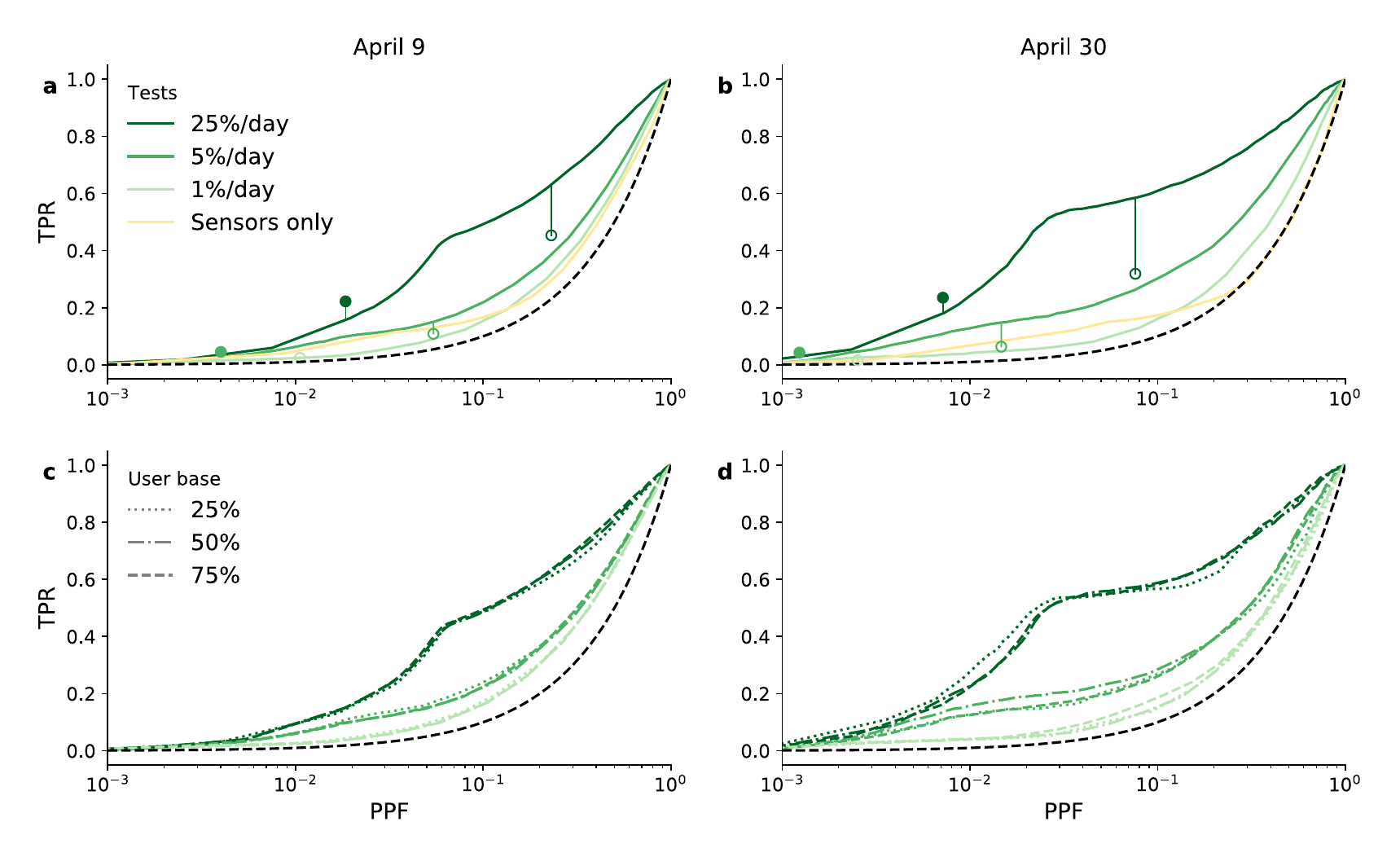}
    
    \end{center}
    \caption{Receiver operating characteristic (ROC) curves for classification as possibly infectious. As in Fig~\ref{f:ROC_curves}, but for subnetworks with randomly selected nodes rather than for subnetworks with a neighborhood topology. For the filled circles, the 1\%/day case falls outside the plotting region; values for panel (a) are (7$\times10^{-4}$, 0.009) and for panel (b) are (2$\times10^{-4}$, 0.01).}
    \label{f:ROC_curves_random_subnet}
\end{figure*}

\clearpage

\begin{figure*}
    \begin{center}
    \includegraphics[width=0.95\textwidth]{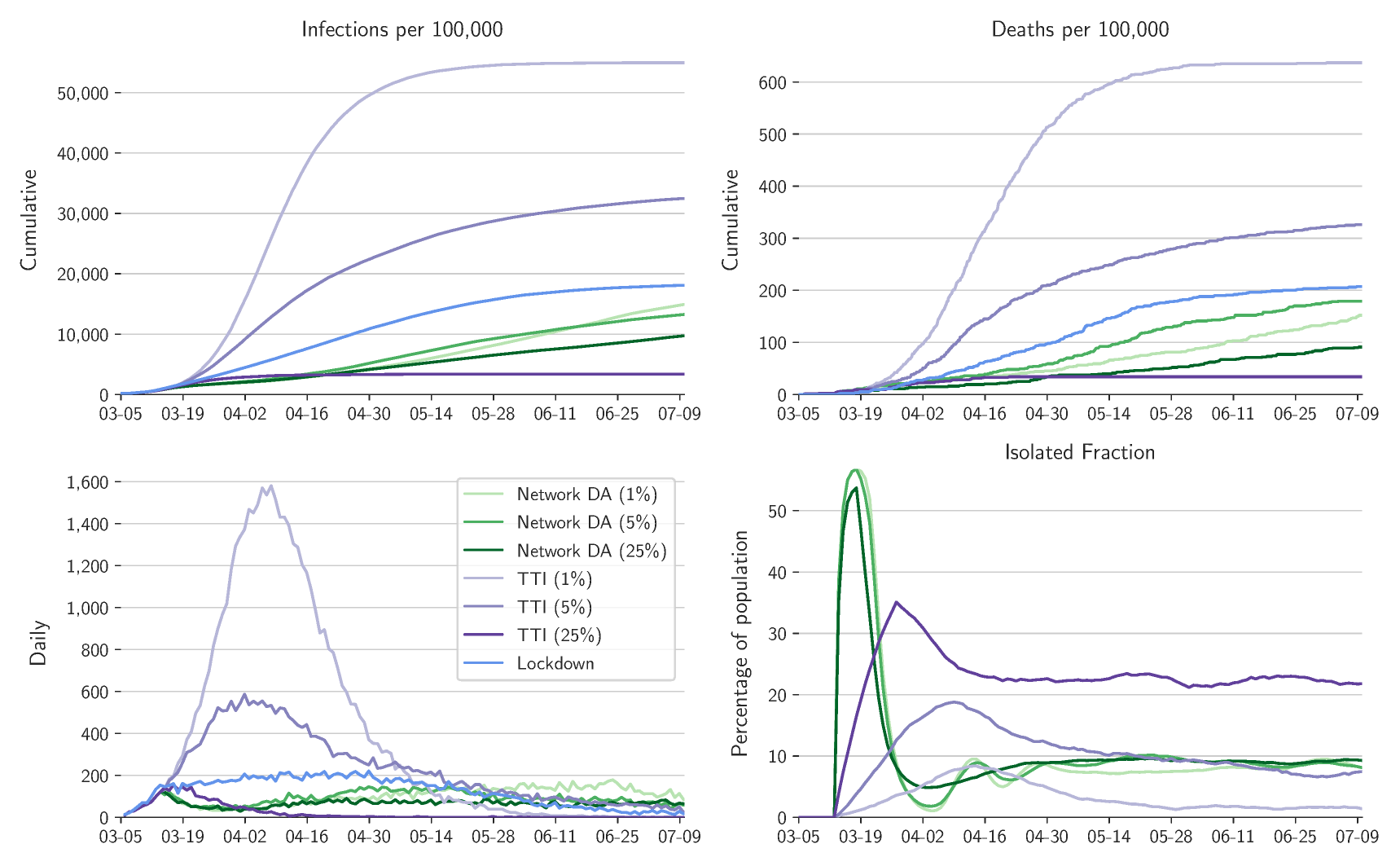}
    \end{center}
    \caption{Comparison of different contact intervention scenarios for random user base with $\tilde N/N=75\%$. As in Fig.~\ref{f:intervention_u75nbhd}, but with a subnetwork with randomly selected nodes and with a classification threshold $c_I = 0.25\%$.}
    \label{f:intervention_u75rand}
\end{figure*}

\clearpage

\begin{figure*}
    \begin{center}
    \includegraphics[width=0.95\textwidth]{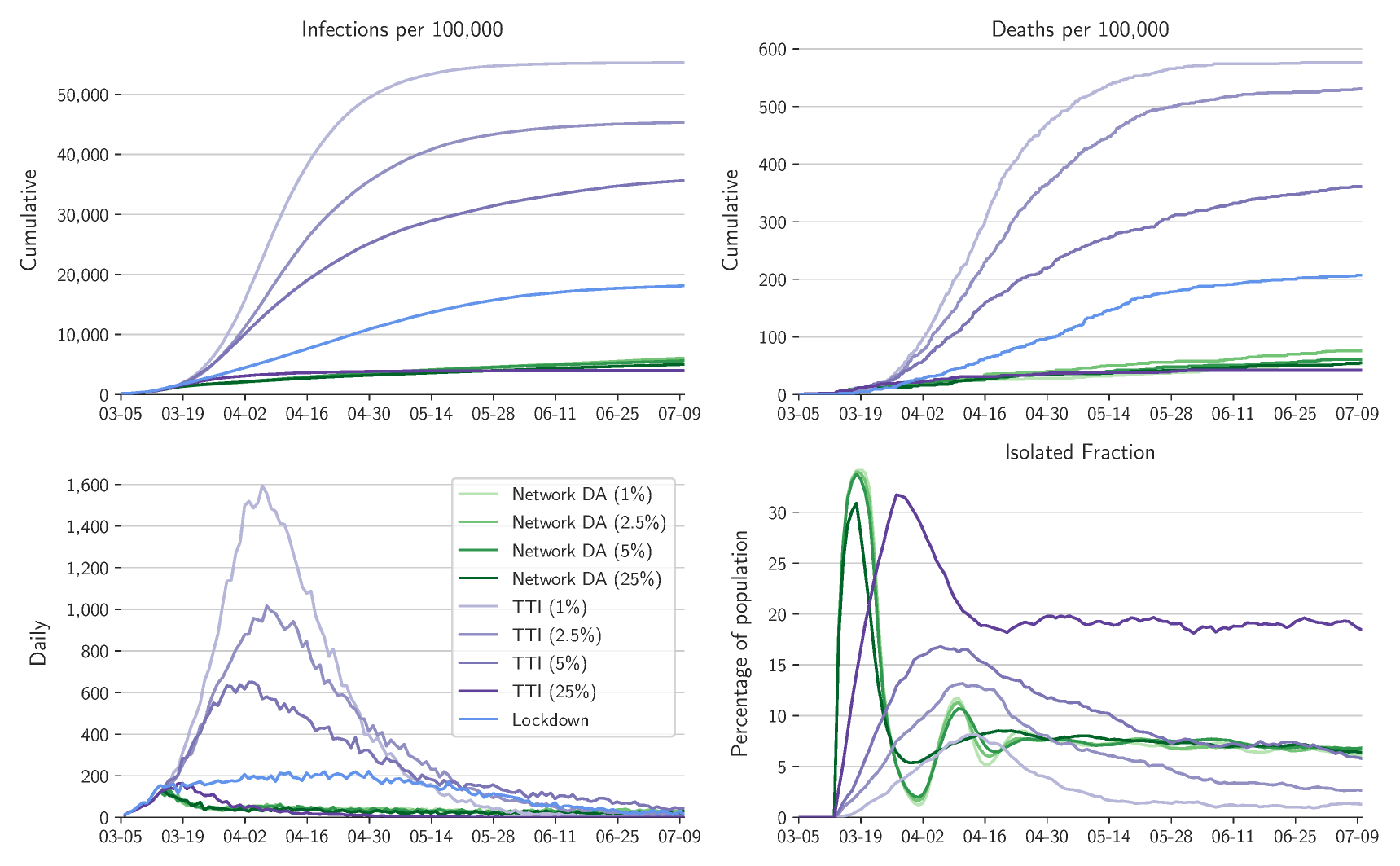}
    \end{center}
    \caption{Comparison of different contact intervention scenarios for neighborhood user base with $\tilde N/N=50\%$ and with a classification threshold $c_I = 0.5\%$.}
    \label{f:intervention_u50nbhd}
\end{figure*}

\clearpage

\begin{figure*}
    \begin{center}
    \includegraphics[width=0.95\textwidth]{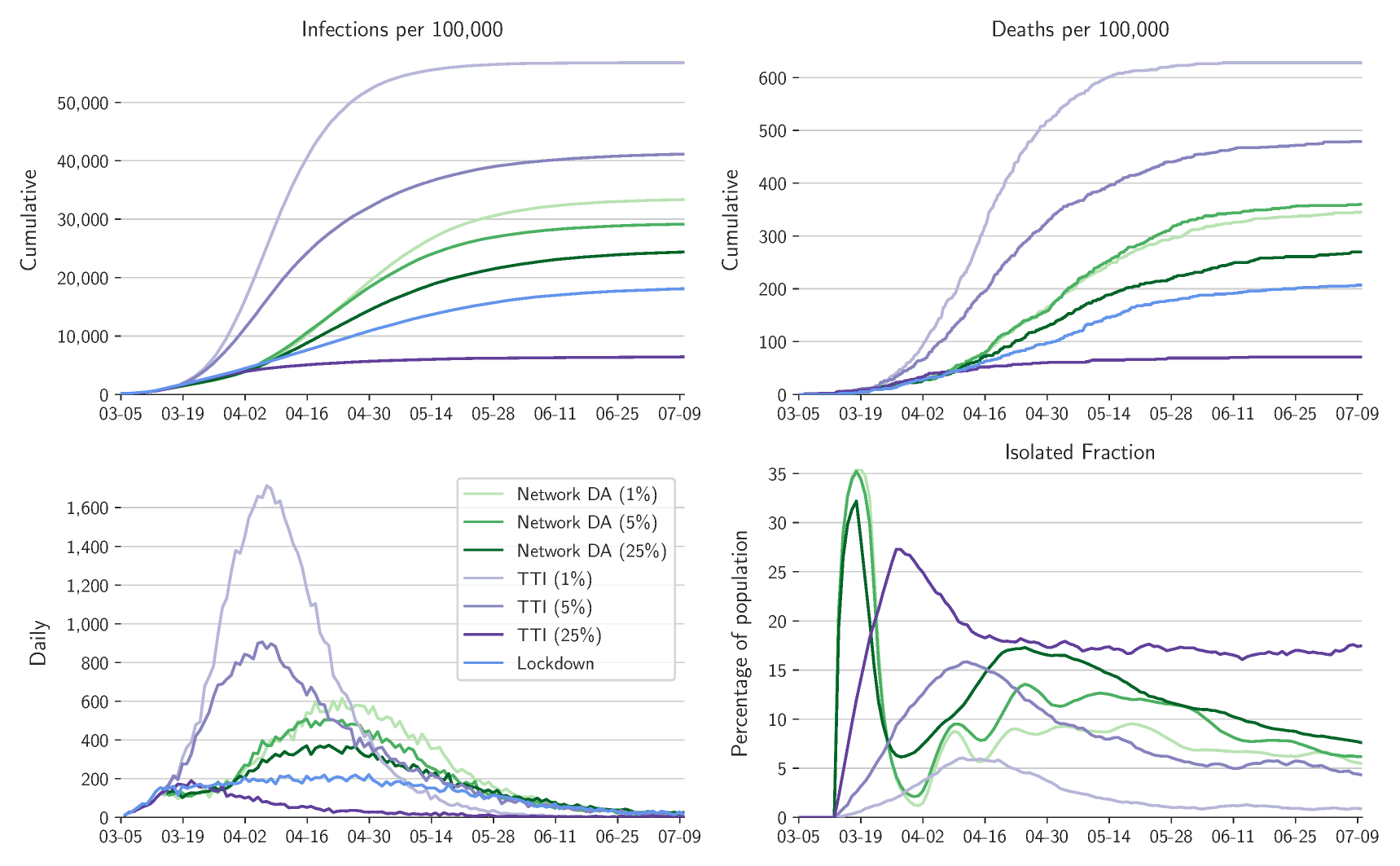}
    \end{center}
    \caption{Comparison of different contact intervention scenarios for random user base with $\tilde N/N=50\%$ and with a lower classification threshold $c_I = 0.25\%$. As in Fig.~\ref{f:intervention_u50nbhd}, but with a subnetwork with randomly selected nodes.}
    \label{f:intervention_u50rand}
\end{figure*}

\clearpage

\begin{figure*}
    \begin{center}
    \includegraphics[width=0.95\textwidth]{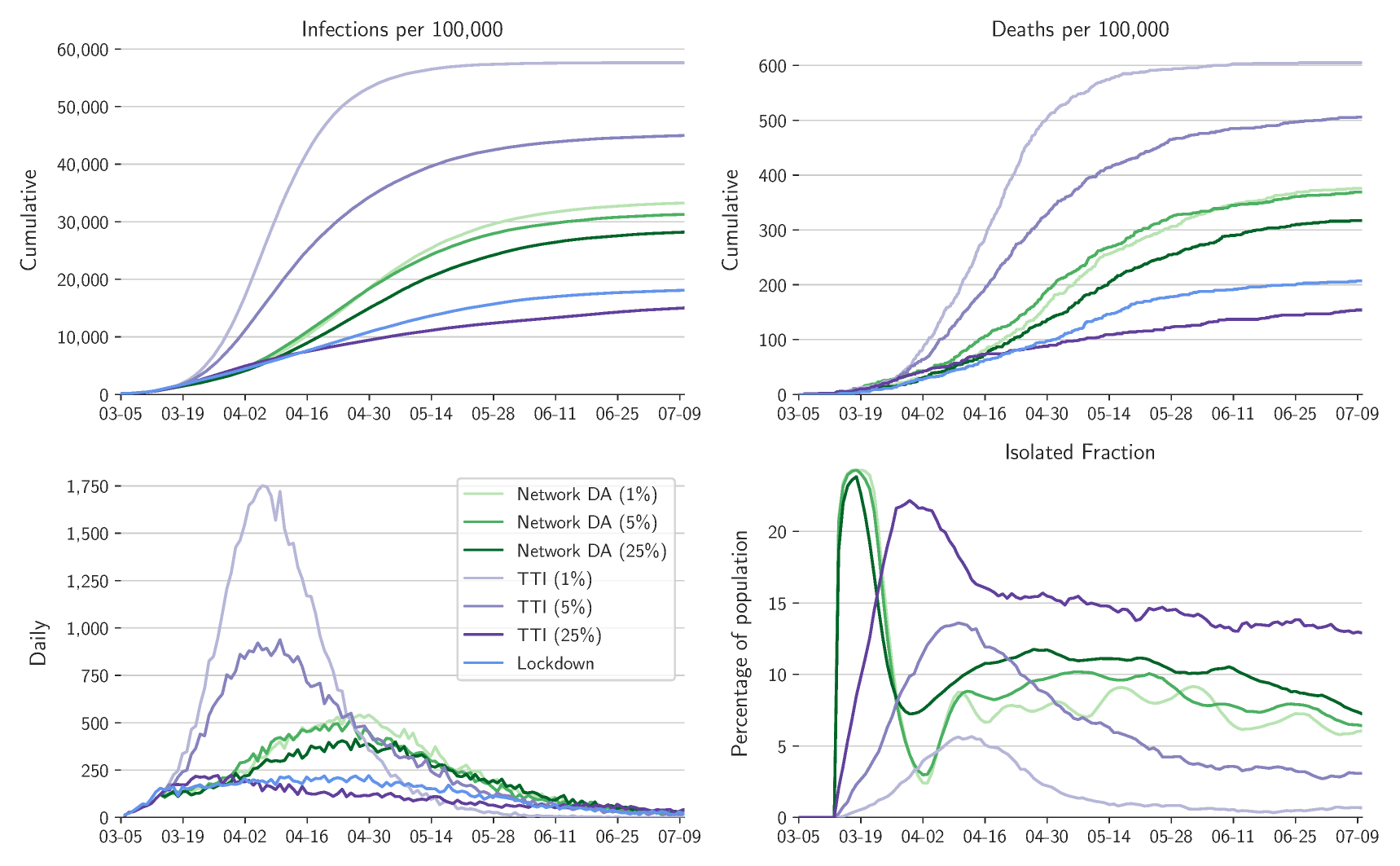}
    \end{center}
    \caption{Comparison of different contact intervention scenarios for neighborhood user base with $\tilde N/N=25\%$ and with a classification threshold $c_I = 0.25\%$.}
    \label{f:intervention_u25nbhd}
\end{figure*}

\clearpage

\begin{figure*}
    \begin{center}
    \includegraphics[width=0.95\textwidth]{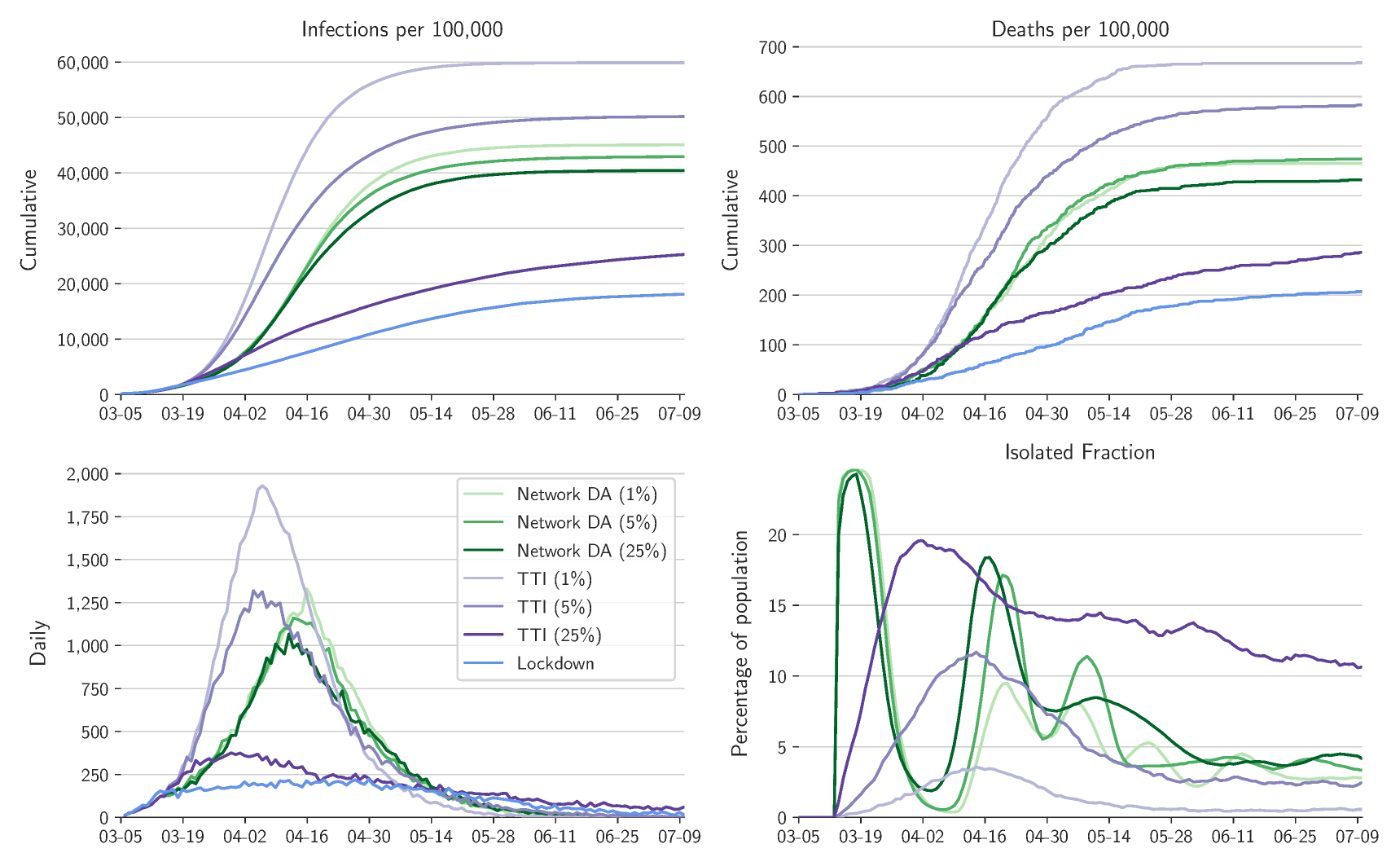}
    \end{center}
    \caption{Comparison of different contact intervention scenarios for random user base with $\tilde N/N=25\%$ and with a lower classification threshold $c_I = 0.01\%$. As in Fig.~\ref{f:intervention_u25nbhd}, but with a subnetwork with randomly selected nodes.}
    \label{f:intervention_u25rand}
\end{figure*}

\clearpage
 
 \begin{figure*}
    \begin{center}
    \includegraphics[width=0.95\textwidth]{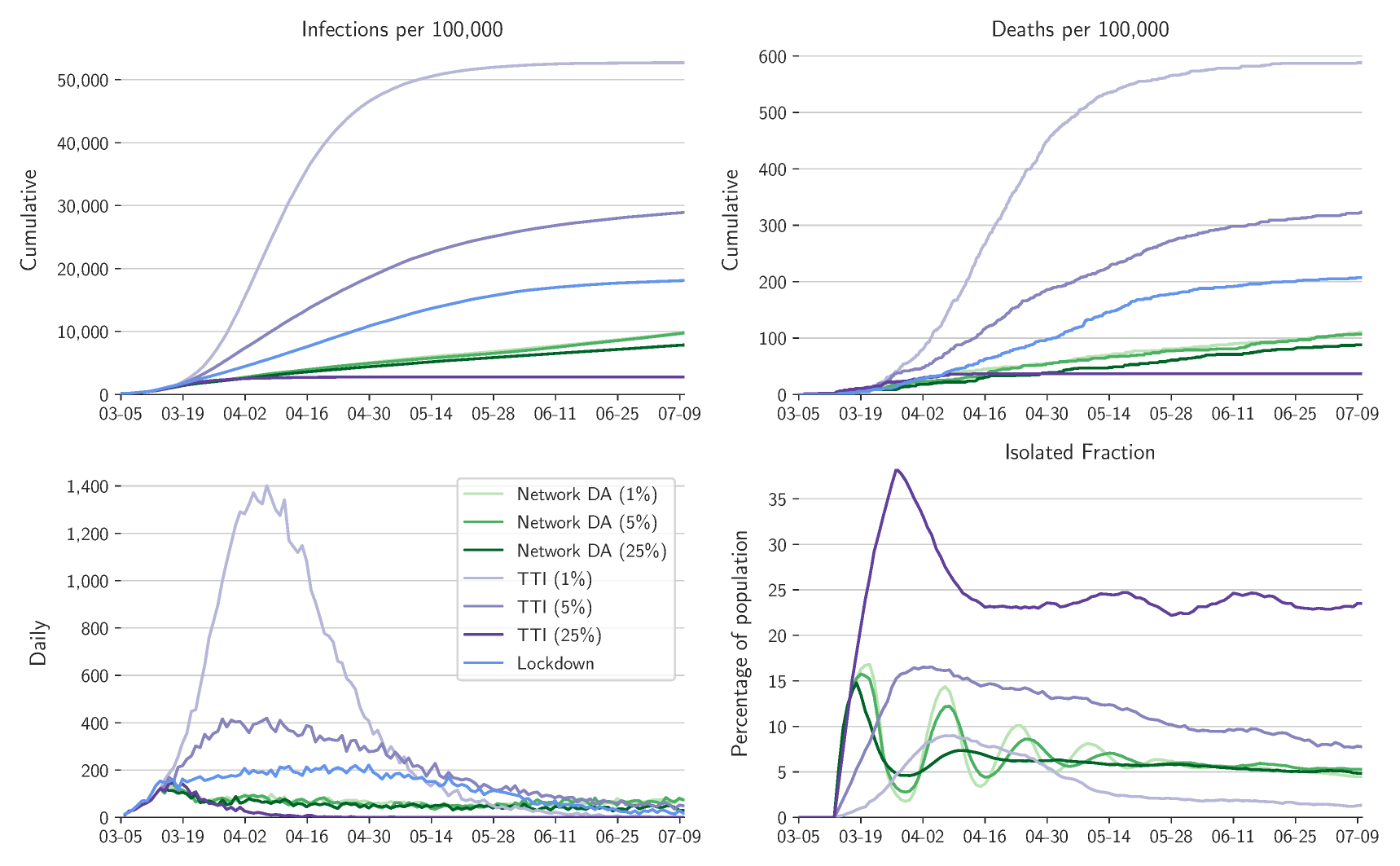}
    \end{center}
    \caption{As in Fig.~\ref{f:intervention_u75nbhd}, comparison of different contact intervention scenarios for neighborhood user base with $\tilde N/N=75\%$ and with a classification threshold $c_I = 1\%$, but replacing the user-dependent number of external neighbours $k^x_i$ by the constant exterior connectivity from Table \ref{tab:user_base}.}
    \label{f:intervention_u75nbhd_avgexternal}
\end{figure*}

\clearpage

\begin{figure*}
    \begin{center}
    \includegraphics[width=0.95\textwidth]{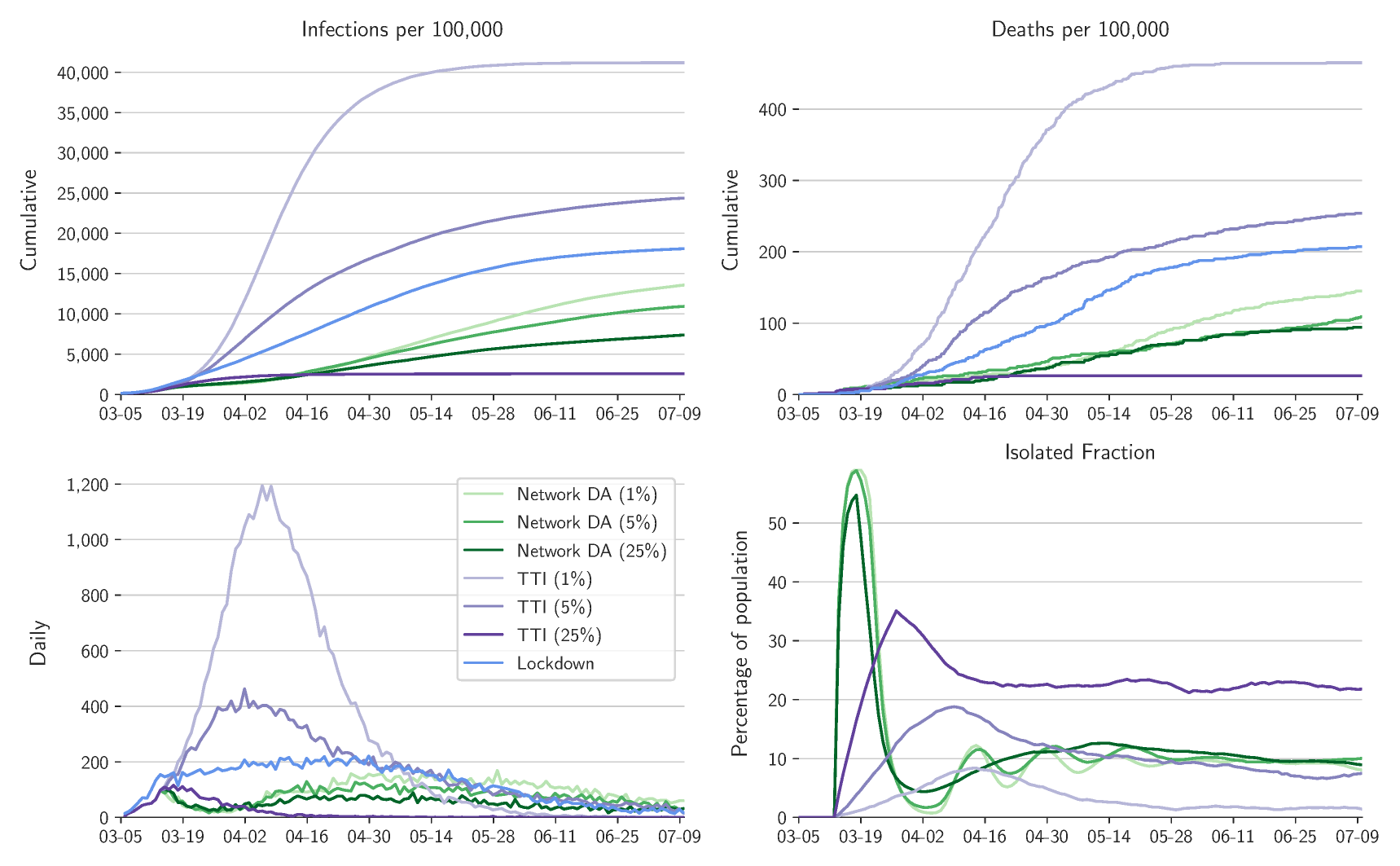}
    \end{center}
    \caption{As in Fig.~\ref{f:intervention_u75rand}, comparison of different contact intervention scenarios for random user base with $\tilde N/N=75\%$ and with a classification threshold $c_I = 0.25\%$, but replacing the user-dependent number of external neighbours $k^x_i$ by the constant exterior connectivity from Table \ref{tab:user_base}.}
    \label{f:intervention_u75rand_avgexternal}
\end{figure*}

  \clearpage

\begin{figure*}
    \begin{center}
    \includegraphics[width=\textwidth]{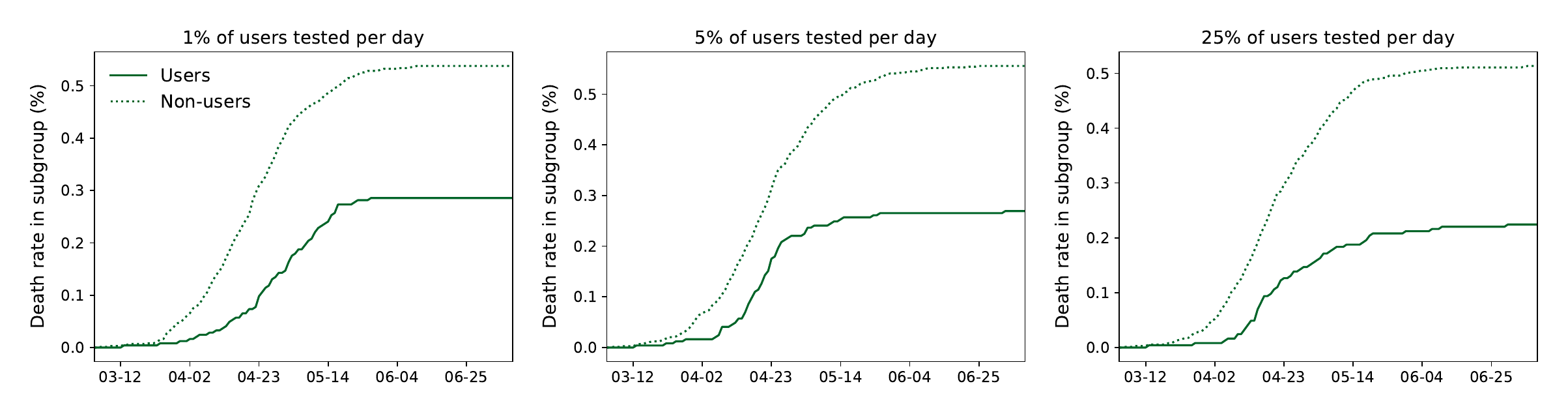}
    \end{center}
    \caption{Cumulative death rate of users vs.\ non-users for the $\tilde{N}/N=25\%$ user base consisting of nodes selected at random from the overall population network. Individual contact interventions are applied within the user base from March 15 onward.}
    \label{f:user_vs_nonuser_rand}
\end{figure*}

\clearpage

\begin{figure*}
    \begin{center}
    \includegraphics[width = \textwidth]{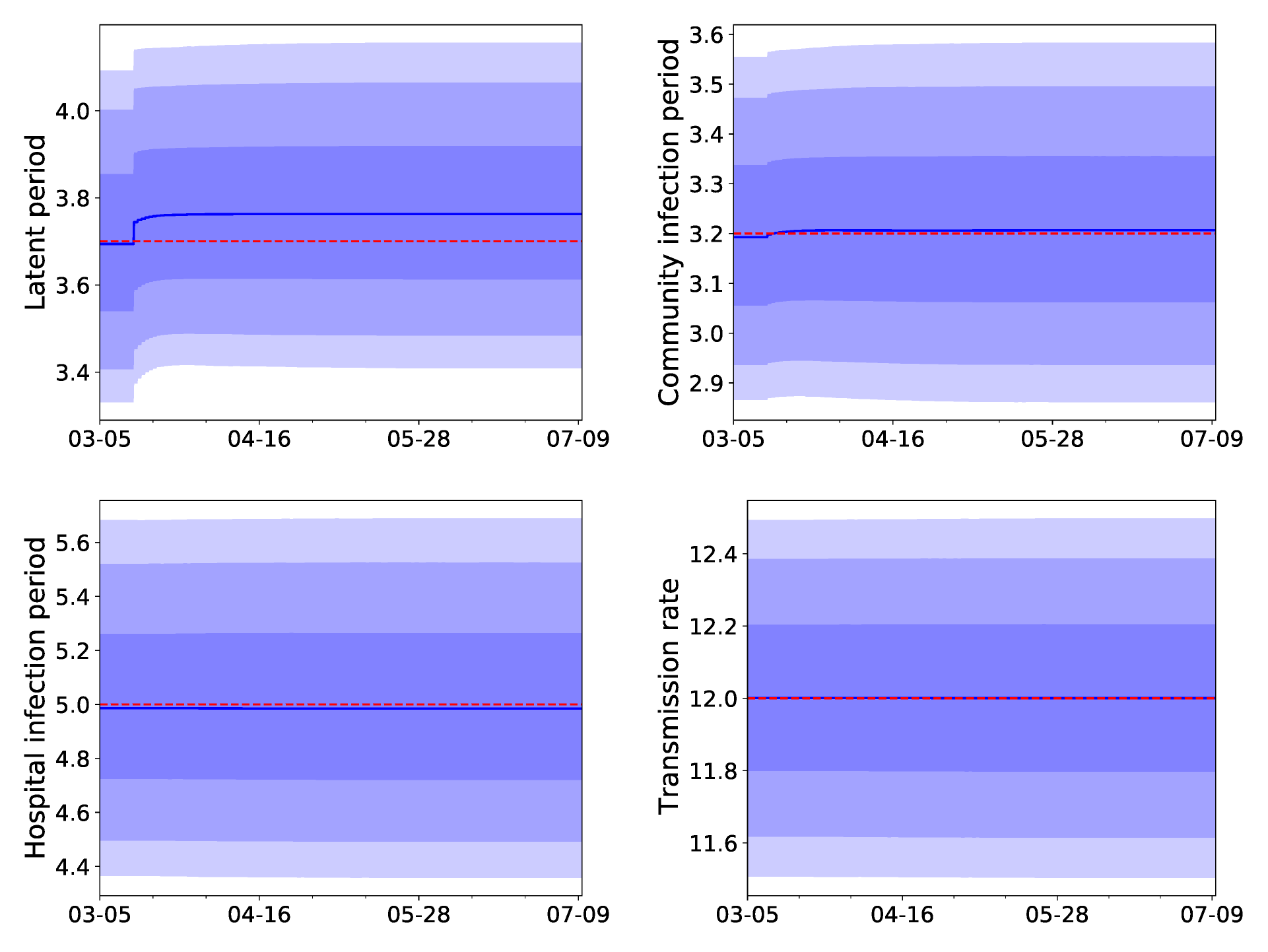}
    \end{center}
    \caption{Distribution of ensemble averaged model parameters across nodes as a function of time during the epidemic. The shaded regions contain 50\%, 80\% and 90\% of the distribution. The dashed line represents the true parameters in the stochastic simulation. During the first 8 days, no DA is performed, and the parameter distributions are the prior distributions.}    
    \label{fig:param_posterior}
\end{figure*}

\end{document}